\documentclass[superscriptaddress,reprint,amsmath,amssymb,aps,floatfix]{revtex4-2}
\usepackage{xcolor}
\usepackage{graphicx}% Include figure files
\usepackage{dcolumn}% Align table columns on decimal point
\usepackage{bm}% bold math
\usepackage{hyperref}% add hypertext capabilities
\usepackage[mathlines]{lineno}% Enable numbering of text and display math
\usepackage{verbatim}

\begin{document}
\preprint{APS/123-QED}
%\preprint{Phys. Rev. Lett.}

%\title{Finite temperature electronic properties of diamond and diamond-like materials:\\ effects of quantum nuclear motion and disorder}% Force line breaks with \\

\title{Quantum vibronic effects on the electronic properties of solid and molecular carbon}% Force line breaks with \\

\author{Arpan Kundu}
\email[Corresponding author. Email: ]{arpank@uchicago.edu}
\affiliation{Pritzker School of Molecular Engineering, University of Chicago, Chicago, Illinois 60637, United States}

\author{Marco Govoni}
\affiliation{ Materials Science Division and Center for Molecular Engineering, Argonne National Laboratory, Lemont, Illinois 60439, United States.}
\affiliation{Pritzker School of Molecular Engineering, University of Chicago, Chicago, Illinois 60637, United States}

\author {Han Yang}
\affiliation{Department of Chemistry, University of Chicago, Chicago, Illinois 60637, United States}

\author{Michele Ceriotti}
\affiliation{Laboratory of Computational Science and Modeling, IMX, Ecole Polytechnique Federale de Lausanne, 1015 Lausanne, Switzerland}

\author{Francois Gygi}
\affiliation{Department of Computer Science, University of California Davis, Davis, California 95616, United States}

\author{Giulia Galli}
\email[Corresponding author. Email: ]{gagalli@uchicago.edu}
 \affiliation{Pritzker School of Molecular Engineering, University of Chicago, Chicago, Illinois 60637, United States}
 \affiliation{ Materials Science Division and Center for Molecular Engineering, Argonne National Laboratory, Lemont, Illinois 60439, United States.}
\affiliation{Department of Chemistry, University of Chicago, Chicago, Illinois 60637, United States}

\date{\today}% It is always \today, today,

%TC:ignore
\begin{abstract}
We study the effect of quantum vibronic coupling on the electronic properties of carbon allotropes,  including molecules and solids, by combining path integral first principles molecular dynamics (FPMD) with a colored noise thermostat. In addition to avoiding several approximations commonly adopted in calculations of electron-phonon coupling, our approach only adds a moderate computational cost to  FPMD simulations and hence it is applicable to large supercells, such as those required to describe amorphous solids. We predict the effect of electron-phonon coupling on the fundamental gap of amorphous carbon, and we  show that in diamond the  zero-phonon renormalization of the band gap is larger than previously reported.

\begin{description}

\item[Keywords]
Path Integral, Electron-phonon, ZPR, Bandgap

\end{description}
\end{abstract}

\maketitle
%TC:endignore

%\tableofcontents

\newpage

%\input{introduction}
%\section{Introduction}
Understanding the electronic structure of materials and molecules at finite temperature is important for the prediction of the physical properties of countless systems, ranging from  opto- and bio-electronic devices, to solar cells, and materials used to build quantum sensors and quantum computers. However, a general theoretical framework to study the electronic properties of molecules and solids over a wide range of temperatures, incorporating accurately nuclear quantum effects and electron-phonon interaction, is still missing.

Most theoretical studies of electron-phonon coupling have been based either on first principles molecular dynamics (FPMD) \cite{Marx_Hutter_book,Prasanna_preprint_2009, aC_Giulia_PRL_1989,Prasai_rev_2016} or on perturbative calculations assuming harmonic potential energy surface (PES) \cite{el_ph_rev_Cardona_2005,Giustino_Rev_2017,Monserrat_Rev_2018}.
%, with the latter applicable only to ordered systems. 
FPMD is accurate above the Debye temperature, provided inter-atomic interactions are described at an appropriate level of density functional theory. However, for light systems, especially those containing first row elements, FPMD may not be appropriate, since nuclear quantum effects play an important role even at ambient conditions. Notable examples are liquid water \cite{Ceriotti_Chem_Rev_2016} and ice \cite{Pamuk_PRL_2012,Buxton_JCP_2019}, many molecular crystals \cite{Rossi_PRL_2016,Cazorla_RevModPhys_2017}, and materials and molecules composed mostly of carbon atoms, such as polymers, diamond, and graphite. In principle, perturbative \cite{el_ph_rev_Cardona_2005,Giustino_Rev_2017,Monserrat_Rev_2018} and non-perturbative stochastic \cite{Zacharias_PRL_2015,Zacharias_PRB_2016,Stoch_Kresse_NJP2018} approaches, with anharmonic effects included at various levels of approximation \cite{Antonius_PRB_2015,VSCF_Monserrat_2013}, may be used also below the Debye temperature, and they have been applied to several crystalline solids \cite{Zacharias_PRL_2015,Zacharias_PRB_2016,Antonius_PRB_2015,Stoch_Kresse_NJP2018,VSCF_Monserrat_2013,Giustino_PRL_2010,Canuccia_PRL_2011}. However, they are not well suited to study disordered systems, for example amorphous or glassy materials, molecular compounds and nanostructures \cite{Kapil_JCTC_2019}.

Here we investigate the effect of electron-phonon interaction on the electronic properties of solids and molecules by accurately including quantum vibronic effects in first principles simulations.   We used path integral (PI) molecular dynamics with a  colored noise generalized Langevin equation (GLE), named PIGLET, to sample the appropriate quantum fluctuations of the nuclei \cite{PIGLET_Ceriotti_PRL_2012}. In addition, we performed FPMD simulations with a single bead and colored noise GLE (a so-called quantum thermostat (QT) \cite{QT_Ceriotti_PRL_2009}). See section S2 in SI for more details.  We show that the ability to perform PI simulations at a cost comparable to that of FPMD is critical to obtain accurate results for large systems.  %with a computational cost which is an order of magnitude smaller than that of straightforward path integral simulations.  
We report results for several carbon systems, including diamond, amorphous carbon (a-C), and pentamantane and we propose a simple computational protocol to predict the fundamental gap of light disordered solids including nuclear quantum effects (NQE),  in an accurate and efficient manner.   We predict for the first time the effect of electron-phonon coupling on the electronic properties of diamond-like a-C  and we show that the zero-phonon renormalization (ZPR) of the band gap of crystalline diamond is larger than previously reported, due to vibrational anharmonic effects. The approach proposed here permits to assess the validity of commonly used approximations in the calculation of electron-phonon interaction in molecular and condensed systems.

\begin{figure*}[htbp]
\includegraphics{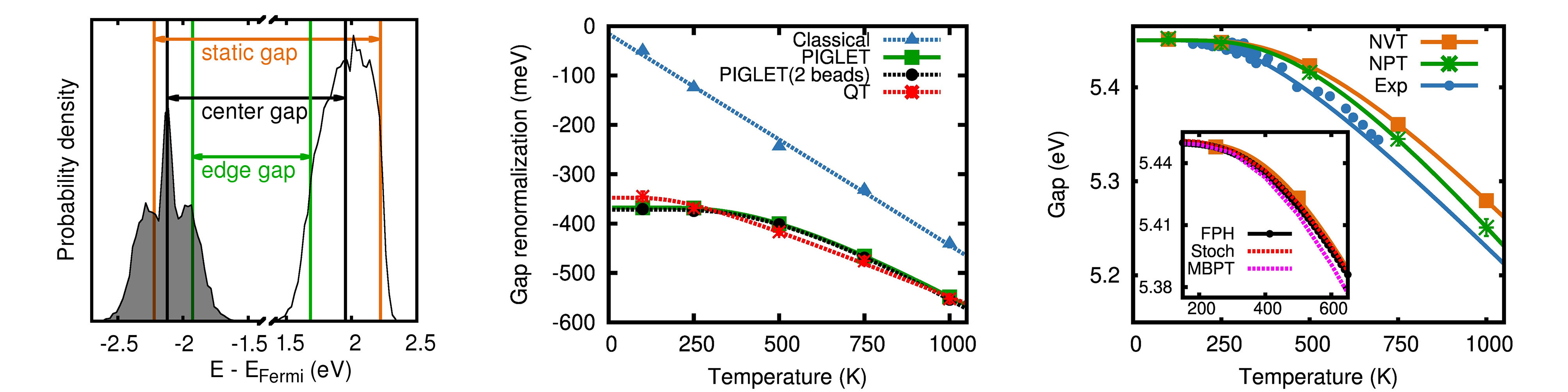}
\caption{(Left panel) Electronic density of states (EDOS) at the valence band maximum (VBM, shaded) and conduction band minimum (CBM) of diamond, computed with a 64 atom supercell and a 16 bead PIGLET NVT simulation at 100 K. The green, black and orange vertical lines represent the thermal average of the band edges (edge gap), the thermal average of three highest VB and six lowest CB eigenvalues (center gap), and the energy of degenerate VBM and CBM computed at the equilibrium geometry (static gap), respectively. (Middle panel) Difference between the static and center gap (gap renormalization) as a function of temperature, obtained with the same protocol used in the left panel.  The symbols represent the simulation results while the lines are the Vi\~{n}a (linear) model fit \cite{Vina_PRB_1984} of the quantum (classical) results, respectively. Classical values are obtained with first principles molecular dynamics; results including nuclear quantum effects are obtained with PIGLET simulations (2 or more beads) and a quantum thermostat (QT).
%If  not visible, simulation error bars are smaller than the symbol size.
(Right panel) Center gap computed with a 216 atom supercell and 2-beads PIGLET simulations in the NVT and NPT ensembles (solid lines: Vi\~{n}a model fit), compared with experimental results\cite{dia_exp_gap_measure,dia_exp_gap_vina_fit}.
All calculated results have been offset by different amounts so that at T=0 they match the experimental bandgap extrapolated to 0 K. The inset shows the differences between results obtained using the frozen phonon harmonic approximation (FPH), a stochastic approach (Stoch) for a C\textsubscript{250} supercell\cite{Stoch_Kresse_NJP2018}, and  many-body perturbation theory (MBPT) calculations performed on a \(4\times4\times4\) q-point grid\cite{DFPT_Han}.
%Indirect band gaps calculated using the stochastic one-shot method for a C\textsubscript{250} supercell are taken from ref. \cite{Stoch_Kresse_NJP2018}. 
Except for the NPT simulations, we used the static lattice parameter (3.568 \AA) for all calculations. 
%obtained by minimizing the stress tensor for the minimum energy structure. 
}
\label{fig:dia_gap}
\end{figure*}

We start by discussing our results for diamond.  At T=0, if we neglect the zero-point motion of the atoms and electron-phonon coupling, the valence band maximum (VBM) and conduction band minimum (CBM) are 3 and 6 fold degenerate, respectively. At finite T, the band edge degeneracies are broken, as shown in Fig. \ref{fig:dia_gap}, where we  report the electronic density of states (EDOS)  close to the VBM and CBM  of a 64 atom diamond supercell (C\textsubscript{64}), obtained at 100 K from a 16-bead PIGLET simulation.  The renormalized bandgap due to electron-phonon coupling may be defined in two different ways: as the energy difference between  (i) the thermal average of the three eigenvalues associated to the VBM and of the six associated to the CBM  (center gap)\  \cite{PI_herrero_2006,Zacharias_PRL_2015,Zacharias_PRB_2016, Stoch_Kresse_NJP2018,Zacharias_PRR_2020} or between (ii) the thermal average of the highest  of the three VBM eigenvalues and of the lowest of the six CBM eigenvalues (edge gap). We show below that there is a substantial difference of $\simeq$ 160 meV between the center and edge band gaps due to quantum vibronic effects; the minimum (indirect) gap of diamond corresponds to the edge gap.

We first discuss the center gap. The middle panel of Fig. \ref{fig:dia_gap} shows the bandgap renormalization as a function of temperature obtained with different approximations.  Compared to PIGLET results, classical simulations underestimate the bandgap renormalization by more than 200 meV for T$< 500$ K, while a QT accurately captures the NQE. The systematic error present at low T in the QT results, due to the so called zero-point energy leakage \cite{QT_Ceriotti_PRL_2009}, leads to an extrapolated ZPR that is accurate within  30 meV. Interestingly, we found that such systematic error may be reduced to 10 meV by performing PIGLET simulations with just 2 beads, which only require twice the computational cost of a QT simulation.
On the basis of this result, the NQE of all diamond supercells with more than 64 atoms were simulated  using the 2-beads PIGLET protocol in both the canonical and isothermal-isobaric ensembles (labeled NVT and NPT, respectively).
%
%This is an important result regarding the efficiency of our calculations and for all diamond supercells with more than 64 atoms, discussed below, we carried out 2-bead PIGLET simulations in both the canonical and isothermal-isobaric ensembles (labeled ``NVT" and ``NPT", respectively).

The right panel of Fig. \ref{fig:dia_gap} compares the band gaps for a 216 atom supercell (C\textsubscript{216}) with the measured indirect band gap of diamond \cite{dia_exp_gap_measure} fitted using the Vi\~{n}a model \cite{dia_exp_gap_vina_fit}. 
The difference between the bandgaps obtained at constant volume and constant pressure is negligible at low temperature and  it is only 30 meV at 1000 K. Therefore, we conclude that the lattice thermal expansion of diamond has a negligible effect on bandgap calculations, consistent with previous studies \cite{PI_herrero_2006,VSCF_Monserrat_2013} carried out with an approximate treatment of anharmonicity of the PES.

We found a remarkable agreement between results obtained with the stochastic approach \cite{Stoch_Kresse_NJP2018},   frozen phonon harmonic results (FPH, see section S5 in the SI) and NVT simulations (see inset of Fig. \ref{fig:dia_gap}, right panel). This indicates that the anharmonicity of the PES and higher-order electron-phonon couplings have a negligible effect in determining the value of the center gap. In Fig. 1, we also report the results obtained by applying many-body perturbation theory (MBPT) to electron-phonon interactions and using a  generalization of the method of Ref. \cite{DFPT_Ryan} for solids \cite{DFPT_Han}. The method relies on the rigid ion approximation, which assumes that the ionic Hamiltonian depends on potentials created {\it independently} by each nucleus.  The negligible differences between MBPT values and the FPH results indicate that below $\simeq$  500 K the rigid ion approximation is justified in the case of diamond, consistent with the conclusion of Ref. \cite{Antonius_PRB_2015}.  

%(see Fig. 4 of ref. \cite{Antonius_PRB_2015}). 
%
%The results of NVT simulations, ``FPH" calculations ( see section S6 in the SI) the stochastic method of ref. \cite{Stoch_Kresse_NJP2018} and ``DFPT+MBPT" approach of ref. \cite{DFPT_Han} are comparable at low T. However the FPH results are considerably different above 500K, where high-order \((>2)\) electron-phonon couplings become important and consequently, the quadratic approximation of Kohn-Sham eigenvalues is no longer adequate, as shown by Antonius \textit{et al} (see Fig. 4 of ref. \cite{Antonius_PRB_2015}). 
Furthermore the VSCF calculations by Monserrat \textit{et al.} yielded a ZPR (-462 meV) \cite{VSCF_Monserrat_2013} value which is considerably larger (by about 30 \%) than  the corresponding harmonic (-325 meV) \cite{Thermal_lines_Monserrat},  and our FPH (-321 meV) values, as well as higher than the NVT-PIGLET result (at 100 K, -325 meV), where all calculations were performed for the same supercell (C\textsubscript{250}).
%This indicates that the VSCF approach is not adequate to incorporate anharmonicity in  electron-phonon calculations.
 In contrast, by sampling the PES along each phonon mode and  using the independent mode approximation,  Antonius et al. \cite{Antonius_PRB_2015} found that anharmonicity reduces the ZPR of the {\it direct} band gap of diamond by 30\%. We note however that large displacements (up to 0.3 \AA) along  phonon modes were used in Ref. \cite{Antonius_PRB_2015}  which may introduce a fictitious coupling between (i) stretching and (ii) bending or torsional modes \cite{Piccini_1,Piccini_2} and hence introduce numerical artifacts.

\begin{figure}[htbp]
\includegraphics{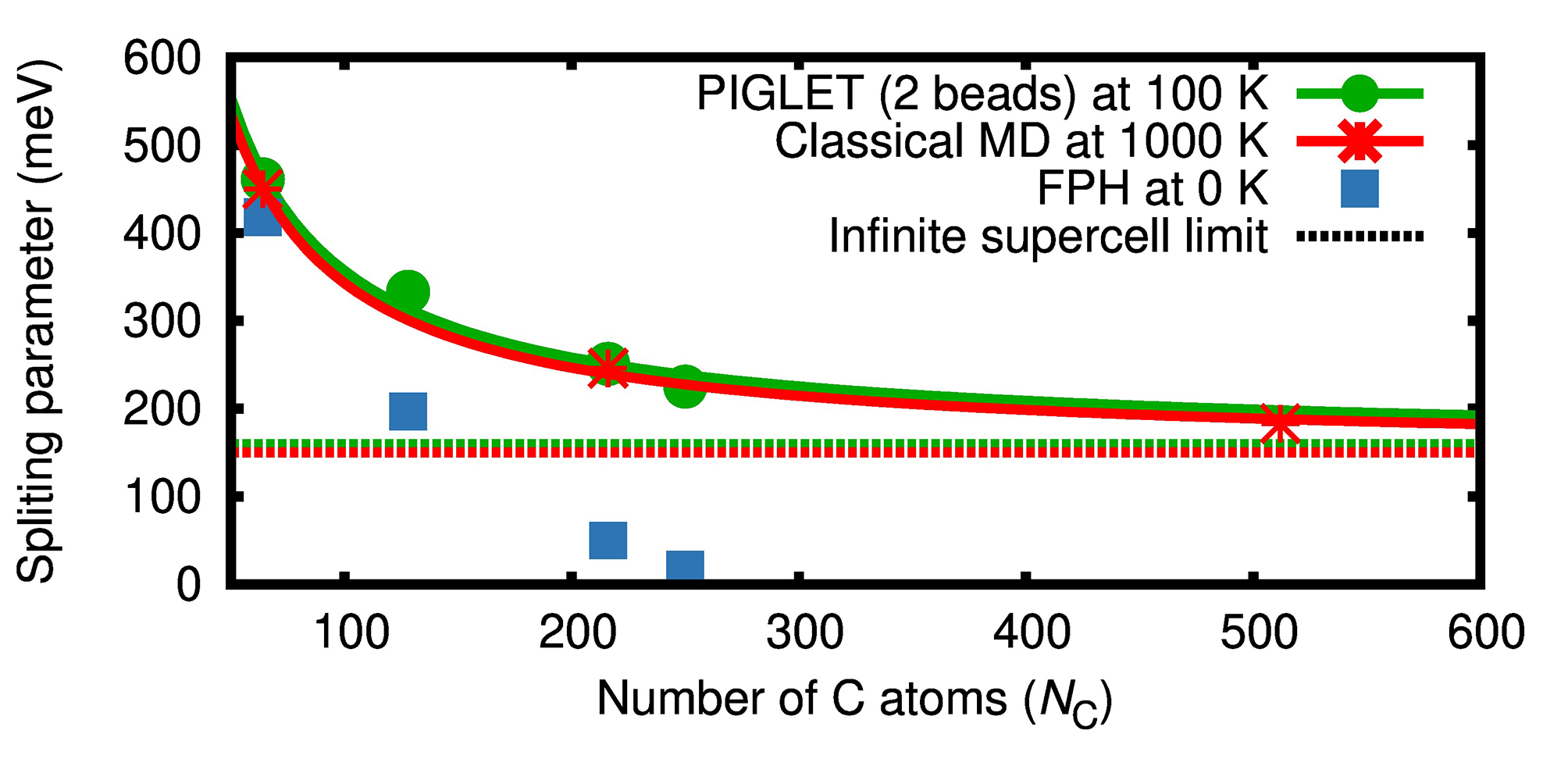}
\caption{Difference between the center and edge gaps of diamond (splitting parameter, which is positive  definite) as a function of the number of atoms \((N_\text{C})\) in the supercell used in our calculations. Symbols represent the simulated data (same acronyms as in Fig.1), which were fitted (solid lines) with the function \(a+b/N_\text{C}\). The dashed lines represent the extrapolated value to the infinite supercell limit.}
\label{fig:dia_extrapol}
\end{figure}

%\begin{figure}[htbp]
%\includegraphics{dia-split-extrapol_lin_v4.jpg}
%\caption{Difference between the center and edge gaps of diamond (splitting parameter, which is positive  definite) as a function of the reciprocal of the number of atoms \((N_\text{C})\) in the supercell used in our calculations. Symbols represent the simulated data (same acronyms as in Fig.1), which were fitted (solid lines) with the function \(a+b/N_\text{C}\). The dashed lines represent the extrapolated value to the infinite supercell limit.}
%\label{fig:dia_extrapol}
%\end{figure}

We now turn to discussing the edge gap of diamond,  defined as the difference between the lowest non degenerate eigenvalue and the highest non degenerate eigenvalue. Previous works \cite{Zacharias_PRB_2016,Zacharias_PRR_2020} concluded that the observed splitting of degenerate bands at T close to zero is caused by the small size of the supercell adopted in DFT calculations.
Zacharias and Giustino \cite{Zacharias_PRR_2020} claimed that  the lifting of the degeneracy is caused only by the zone center phonons, and showed that their omission in the electron-phonon calculation leads to degenerate eigenvalues.  In the infinite supercell size limit, the influence of  zone center phonons should vanish and the splitting of degenerate bands should go to zero. 
We extrapolated the difference between the center and the edge gap (called here splitting parameter, which is positive definite) with respect to the number of C atoms $(N_\text{C})$ in the supercell, using the function \(a+b/N_\text{C}\), where \textit{a} is a  contribution independent from supercell size.
Fig. \ref{fig:dia_extrapol} shows  our results for the 2-beads PIGLET simulations at 100 K, classical MD simulations at 1000 K, and FPH calculations at 0 K, in a canonical ensemble using different supercell sizes.
The splitting parameter converges to zero for the FPH calculations, but for the PIGLET or classical MD simulations, it converges to a non-negligible value of 160 meV. Therefore, the splitting of the band edges found here cannot be ascribed entirely to the finite size of the supercell and it represents a physical effect.  Specifically, we attribute the center and edge gaps difference to  anharmonic vibronic effects. 
We note that MD simulations sample the anharmonic PES, which is not necessarily symmetric around its minimum (see e.g, \cite{VSCF_Monserrat_2013})  and consequently, the probability distributions along phonon modes are not Gaussians due to skewness. For example at 100 K for a C\textsubscript{216} diamond supercell, we found that 234 (out of 645) phonon mode distributions deviate from a Gaussian distribution due to skewness (see section S6 in SI).
This asymmetry  and the consequent local dynamical disorder (see section S7 in SI) is obviously more pronounced for large amplitude oscillations and hence clearly visible  at high T, e.g. 1000 K, in classical MD simulations. However, when nuclear quantum effects are considered, the asymmetry is present even at  T close to zero.  We emphasize that it is this asymmetry and its contribution to the ZPR that lead to a significant difference of 160 meV between the edge and central gap.
In its current implementation \cite{Zacharias_PRL_2015,Zacharias_PRB_2016,Stoch_Kresse_NJP2018,Zacharias_PRR_2020}, the stochastic method, in the limit of large supercell, does not account for the center-edge splitting  because it samples the probability distribution of connected harmonic oscillators without accounting for the deviations of the PES from  a harmonic well.  
It would be interesting to explore whether the recently proposed SSCHA approximation \cite{Mauri_PRB_2014,Monacelli_preprint_2020} is sufficiently accurate to account for the splitting observed here; we note that the SSCHA method uses linear combinations of symmetric Gaussian functions to construct the anharmonic vibrational wave functions and hence the wave function for a non-symmetric quartic potential becomes symmetric  \cite{Monacelli_preprint_2020}, unlike the exact one. 

\begin{figure}[htbp]
\includegraphics{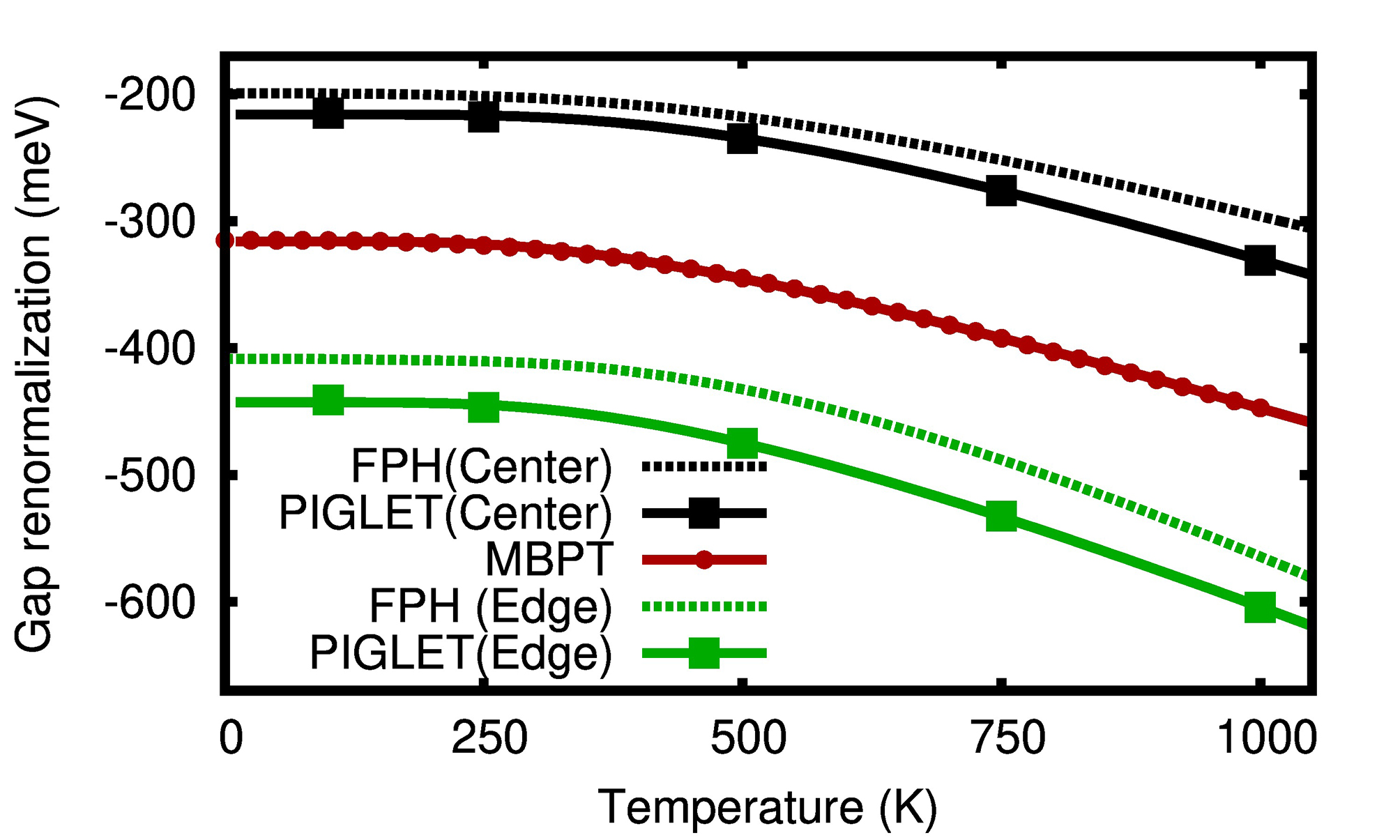}
\caption{
Difference between the static and center and static and edge  fundamental gap  of a pentamantane molecule (gap renormalization),  C\textsubscript{26}H\textsubscript{32} (T\textsubscript{d}), obtained using  PIGLET simulations, and calculated with the frozen phonon harmonic approach (FPH) and many-body perturbation theory (MBPT). The renormalization of both the center and edge gap is reported in the case of FPH and PIGLET calculations. By construction, the difference between center and edge gaps is zero within MBPT.
}
\label{fig:penta}
\end{figure}

In order to explore whether a difference between center and edge gaps is observed also for carbon nanostructures, we studied the electronic properties of the pentamantane molecule. Fig. \ref{fig:penta} shows the results from PIGLET simulations and FPH calculations (See also Fig. S8 in the SI). The ZPR of the center gap  (FPH: -200 meV, PIGLET: -220 meV) is consistent with previous estimates at the PBE level of theory (-210 meV) \cite{FPH_Bester_2019}. However, the ZPR of the  edge gap (FPH: -405 meV, PIGLET: -445 meV) is twice as large and certainly this value is not affected by the size of the supercell, since we are considering an isolated molecule. The presence of the splitting observed here was also reported 
in previous \textit{ab-initio} studies of small diamondoids \cite{Giustino_Nat_Commun_2013,Gali_Nat_Commun_2016}, and Gali \textit{et al.} suggested that the fine structure of the diamondoid photoemission spectra can only be  explained by considering such a splitting \cite{Gali_Nat_Commun_2016}.
We suggest that,  as in the case of crystalline diamond, the edge gap is the most appropriate definition for the  single particle gap in the case of the pentamantane molecule,  since by definition the measurement of a gap by photoemission is a measure of the energy difference between the highest occupied and the lowest unoccupied single particle orbitals.
Note that, unlike diamond, the splitting observed for pentamantane does not only stem from anharmonic effects because FPH calculations also yield a sizeable splitting. 
The comparison of FPH and PIGLET results for pentamantane shows that even at high temperatures, the combined contribution of (i) anharmonicity of the PES and (ii) higher order electron-phonon coupling does not amount to more than 10\% of the total electron-phonon renormalizations of the edge gap. 

In Fig. \ref{fig:penta}, we also show the finite temperature electron-phonon renormalizations calculated using the MBPT approach as implemented in the WEST code\cite{Govoni2015} and presented in Ref. \cite{DFPT_Ryan}.
%considering Fan-Migdal and Debye-Waller electron-phonon self energies and using the Bose-Einstein statistics to represent the temperature dependence (see Eqs. (3-4) in Ref. \cite{DFPT_Ryan}). 
In addition to the harmonic approximation, the calculations of Ref. \cite{DFPT_Ryan} used the rigid-ion approximation, finding a result for the center gap which differs by 100 meV from that obtained with the FPH approach. This difference indicates that the rigid ion approximation is not sufficiently accurate for molecular systems, e.g., isolated molecules, and, we expect, for molecular crystals as well. 

\begin{figure}[htbp]
\includegraphics{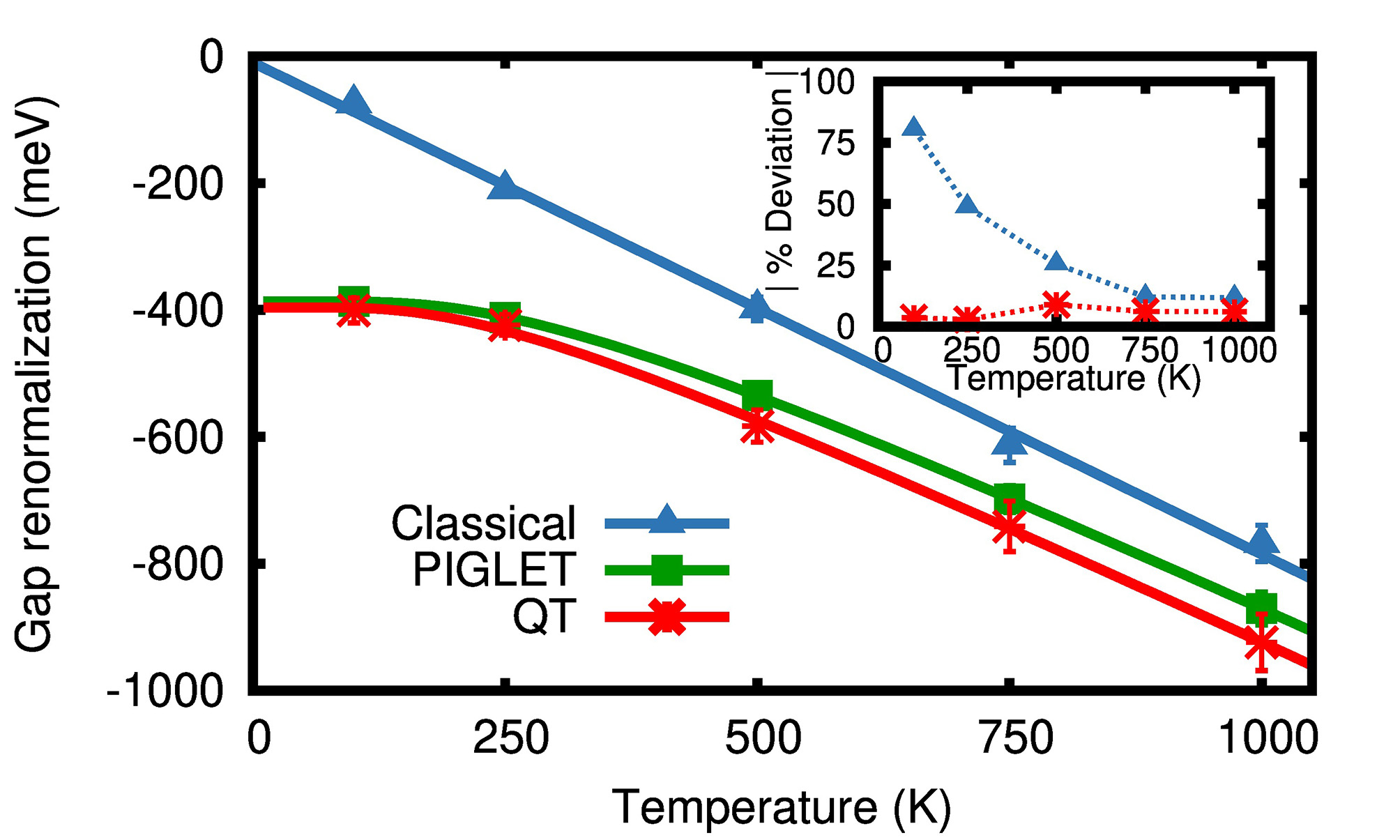}
\caption{
Difference between the minimum static and center gap (gap renormalization) computed  for amorphous-C obtained using different methods to treat nuclear quantum effects (acronyms as in Fig.1). We carried out NVT simulations for an a-C sample with density of 3.25 g/cm\textsuperscript{3}, from  ref. \cite{aC_Risplendi_APL_2014}.
%If not visible, simulation error bars are smaller than the symbol size.
The inset shows the absolute percentage deviation of (i) classical molecular dynamics (MD) and (ii) MD with a quantum thermostat from PIGLET simulations.
}
\label{fig:aC}
\end{figure}

As in the case of the diamond crystal, we compared QT and PIGLET gap renormalization results for pentamantane over a wide range of temperatures and found excellent agreement (see Fig S9), indicating that the use of a QT is adequate for MD simulation studies of single particle gaps of  molecular systems, and it is expected to be particularly valuable  for larger nanostructures and when costly functionals such as meta-GGA or hybrid functionals are adopted.

Finally, we present results for a diamond-like amorphous carbon (DLC), which is an example of a disordered system, where both localized and extended electronic states are present and where there are no degenerate electronic states. In addition to classical MD, we used PIGLET as well as MD with a QT to investigate the effects of NQE on the renormalization of the minimum gap, and the results are shown in Fig \ref{fig:aC} for a 216 atom sample.

For T $<$ 250 K, neglecting NQE severely affects the bandgap renormalization, which is underestimated by more than 50\% in the classical MD simulations, relative to the PIGLET results. As T is increased, as expected, the classical description becomes increasingly more accurate, with only $\sim$10\% deviations at 1000 K. The bandgap predictions from classical MD are, therefore, not accurate for most applications with a working temperature range of 250 - 350 K.
%In order to model a realistic amorphous sample, \textcolor{blue}{it is necessary to average over multiple configurations obtained during dynamical simulations} \textcolor{red}{[The blue text gives me an impression that we are talking about MD simulation of one aC sample, while I think the problem is we have to simulate many aC samples of same density, each located near different minimum. Each MD would sample nearby region ]} and often large samples with several hundreds of atoms are required to represent the medium range order in the system; hence, performing PIGLET simulations with multiple beads may not be computationally feasible. 
Interestingly, we find that in spite of the presence of some localized states in the system, which are expected to be less sensitive to electron-phonon renormalization than extended states \cite{DFPT_Han}, the overall ZPR of the band gap of a-C is substantial ($\simeq 400$ meV at low T), amounting to about two third of that found for crystalline diamond. Finally, we note the  excellent agreement between QT and PIGLET results below 500 K, confirming that MD simulations with either a QT or 2 beads-PIGLET represent a promising, pragmatic choice that does not add a substantial computational overhead to classical FPMD (see section S3). This result is particularly important for the modelling of amorphous systems, where it is usually  necessary to average results over multiple  configurations obtained from separate annealing processes; in addition, large samples with several hundreds of atoms are often required to represent the medium range order in these systems; hence, performing PIGLET simulations with a large number of beads may not be computationally feasible. 

%\input{conclusion}
%\section{Conclusion}

In summary, we investigated the effect of quantum vibronic coupling on the electronic properties of light molecules and solids, including ordered and disordered systems, by coupling  FPMD with a generalized quantum thermostat which accounts for anharmonic effects in the ionic potential energy surface. Our approach avoids all the  approximations commonly made in calculations of electron-phonon coupling, including the rigid-ion and the harmonic approximation. Importantly, it is an efficient approach, which only adds a moderate computational cost to  FPMD simulations and hence it is applicable to large supercells, such as those required to describe amorphous solids. 

We found that in molecular and  solid carbon based materials, nuclear quantum effects significantly alter the electron-phonon band gap renormalizations at temperatures below 500 K.  Our calculations showed  that in diamond, even at temperatures close to zero, the degeneracy of the band edges is lifted due to vibrational, anharmonic effects, and the resulting zero phonon renormalization (ZPR) of the band gap due to electron-phonon interaction is $\sim$160 meV, larger than previously reported at the same level of theory. %this 160 meV is based on the center edge gap difference extrapolated to infinite cell limit.
 With continuing improvement in the resolution of photoemission experiments \cite{ARPES_Iwasawa2018}, we believe our predictions are amenable to experimental validation. The ZPR is substantial also for diamond-like a-C, albeit about 30$\%$ smaller than in crystalline diamond.
Similar to the solid phases, we also observed a large ZPR (445 meV) for the pentamantane molecule.

Finally, our simulations allowed us to asses the validity of common approximations used in the literature to study electron-phonon coupling. We showed that the rigid-ion approximation, widely applied in MBPT based methods, though adequate for  extended solids such as diamond at low temperatures ($<$ 500 K), is severely deficient for molecular systems (e.g., Pentamantane).  We found that  stochastic non-perturbative methods are promising approaches; however, in their current implementation they cannot account for the splitting of degenerate orbitals originating from the dynamical disorder found in diamond, due to its anharmonic potential energy surface. Work is in progress to apply the computational protocol to heterogeneous and disordered systems where both localized and delocalized electronic states are present, for example point defects in diamond and amorphous and glassy carbon with different densities.

%TC:ignore
\begin{acknowledgments}
We thank S. Kundu and B. Monserrat for useful discussions. We thank G. Cicero and F. Risplendi for providing the coordinates of diamond like amorphous carbon simulated from first principles. 
This work was supported by MICCoM, as part of the Computational Materials Sciences Program funded by the U.S. Department of Energy. 
This research used resources of the University of Chicago Research Computing Center.
\end{acknowledgments}

\bibliographystyle{apsrev4-2}
\bibliography{manuscript1.bib}% Produces the bibliography via BibTeX.

\providecommand{\noopsort}[1]{}\providecommand{\singleletter}[1]{#1}%
\begin{thebibliography}{40}%
\makeatletter
\providecommand \@ifxundefined [1]{%
 \@ifx{#1\undefined}
}%
\providecommand \@ifnum [1]{%
 \ifnum #1\expandafter \@firstoftwo
 \else \expandafter \@secondoftwo
 \fi
}%
\providecommand \@ifx [1]{%
 \ifx #1\expandafter \@firstoftwo
 \else \expandafter \@secondoftwo
 \fi
}%
\providecommand \natexlab [1]{#1}%
\providecommand \enquote  [1]{``#1''}%
\providecommand \bibnamefont  [1]{#1}%
\providecommand \bibfnamefont [1]{#1}%
\providecommand \citenamefont [1]{#1}%
\providecommand \href@noop [0]{\@secondoftwo}%
\providecommand \href [0]{\begingroup \@sanitize@url \@href}%
\providecommand \@href[1]{\@@startlink{#1}\@@href}%
\providecommand \@@href[1]{\endgroup#1\@@endlink}%
\providecommand \@sanitize@url [0]{\catcode `\\12\catcode `\$12\catcode
  `\&12\catcode `\#12\catcode `\^12\catcode `\_12\catcode `\%12\relax}%
\providecommand \@@startlink[1]{}%
\providecommand \@@endlink[0]{}%
\providecommand \url  [0]{\begingroup\@sanitize@url \@url }%
\providecommand \@url [1]{\endgroup\@href {#1}{\urlprefix }}%
\providecommand \urlprefix  [0]{URL }%
\providecommand \Eprint [0]{\href }%
\providecommand \doibase [0]{https://doi.org/}%
\providecommand \selectlanguage [0]{\@gobble}%
\providecommand \bibinfo  [0]{\@secondoftwo}%
\providecommand \bibfield  [0]{\@secondoftwo}%
\providecommand \translation [1]{[#1]}%
\providecommand \BibitemOpen [0]{}%
\providecommand \bibitemStop [0]{}%
\providecommand \bibitemNoStop [0]{.\EOS\space}%
\providecommand \EOS [0]{\spacefactor3000\relax}%
\providecommand \BibitemShut  [1]{\csname bibitem#1\endcsname}%
\let\auto@bib@innerbib\@empty
%</preamble>
\bibitem [{\citenamefont {Marx}\ and\ \citenamefont
  {Hutter}(2009)}]{Marx_Hutter_book}%
  \BibitemOpen
  \bibfield  {author} {\bibinfo {author} {\bibfnamefont {D.}~\bibnamefont
  {Marx}}\ and\ \bibinfo {author} {\bibfnamefont {J.}~\bibnamefont {Hutter}},\
  }\href {https://books.google.com/books?id=VRZUw8Wk4CIC} {\emph {\bibinfo
  {title} {Ab Initio Molecular Dynamics: Basic Theory and Advanced Methods}}}\
  (\bibinfo  {publisher} {Cambridge University Press},\ \bibinfo {year}
  {2009})\BibitemShut {NoStop}%
\bibitem [{\citenamefont {Prasanna}(2009)}]{Prasanna_preprint_2009}%
  \BibitemOpen
  \bibfield  {author} {\bibinfo {author} {\bibfnamefont {T.~R.~S.}\
  \bibnamefont {Prasanna}},\ }\href@noop {} {\bibinfo {title} {Relation between
  ab initio molecular dynamics and electron-phonon interaction formalisms}}
  (\bibinfo {year} {2009}),\ \Eprint {https://arxiv.org/abs/0902.0719}
  {arXiv:0902.0719} \BibitemShut {NoStop}%
\bibitem [{\citenamefont {Galli}\ \emph {et~al.}(1989)\citenamefont {Galli},
  \citenamefont {Martin}, \citenamefont {Car},\ and\ \citenamefont
  {Parrinello}}]{aC_Giulia_PRL_1989}%
  \BibitemOpen
  \bibfield  {author} {\bibinfo {author} {\bibfnamefont {G.}~\bibnamefont
  {Galli}}, \bibinfo {author} {\bibfnamefont {R.~M.}\ \bibnamefont {Martin}},
  \bibinfo {author} {\bibfnamefont {R.}~\bibnamefont {Car}},\ and\ \bibinfo
  {author} {\bibfnamefont {M.}~\bibnamefont {Parrinello}},\ }\href
  {https://doi.org/10.1103/PhysRevLett.62.555} {\bibfield  {journal} {\bibinfo
  {journal} {Phys. Rev. Lett.}\ }\textbf {\bibinfo {volume} {62}},\ \bibinfo
  {pages} {555} (\bibinfo {year} {1989})}\BibitemShut {NoStop}%
\bibitem [{\citenamefont {Prasai}\ \emph {et~al.}(2016)\citenamefont {Prasai},
  \citenamefont {Biswas},\ and\ \citenamefont {Drabold}}]{Prasai_rev_2016}%
  \BibitemOpen
  \bibfield  {author} {\bibinfo {author} {\bibfnamefont {K.}~\bibnamefont
  {Prasai}}, \bibinfo {author} {\bibfnamefont {P.}~\bibnamefont {Biswas}},\
  and\ \bibinfo {author} {\bibfnamefont {D.~A.}\ \bibnamefont {Drabold}},\
  }\href {https://doi.org/10.1088/0268-1242/31/7/073002} {\bibfield  {journal}
  {\bibinfo  {journal} {Semicond Sci Technol}\ }\textbf {\bibinfo {volume}
  {31}},\ \bibinfo {pages} {073002} (\bibinfo {year} {2016})}\BibitemShut
  {NoStop}%
\bibitem [{\citenamefont {Cardona}(2005)}]{el_ph_rev_Cardona_2005}%
  \BibitemOpen
  \bibfield  {author} {\bibinfo {author} {\bibfnamefont {M.}~\bibnamefont
  {Cardona}},\ }\href
  {https://doi.org/https://doi.org/10.1016/j.ssc.2004.10.028} {\bibfield
  {journal} {\bibinfo  {journal} {Solid State Commun}\ }\textbf {\bibinfo
  {volume} {133}},\ \bibinfo {pages} {3 } (\bibinfo {year} {2005})}\BibitemShut
  {NoStop}%
\bibitem [{\citenamefont {Giustino}(2017)}]{Giustino_Rev_2017}%
  \BibitemOpen
  \bibfield  {author} {\bibinfo {author} {\bibfnamefont {F.}~\bibnamefont
  {Giustino}},\ }\href {https://doi.org/10.1103/RevModPhys.89.015003}
  {\bibfield  {journal} {\bibinfo  {journal} {Rev. Mod. Phys.}\ }\textbf
  {\bibinfo {volume} {89}},\ \bibinfo {pages} {015003} (\bibinfo {year}
  {2017})}\BibitemShut {NoStop}%
\bibitem [{\citenamefont {Monserrat}(2018)}]{Monserrat_Rev_2018}%
  \BibitemOpen
  \bibfield  {author} {\bibinfo {author} {\bibfnamefont {B.}~\bibnamefont
  {Monserrat}},\ }\href {https://doi.org/10.1088/1361-648x/aaa737} {\bibfield
  {journal} {\bibinfo  {journal} {J. Phys. Condens. Matter}\ }\textbf {\bibinfo
  {volume} {30}},\ \bibinfo {pages} {083001} (\bibinfo {year}
  {2018})}\BibitemShut {NoStop}%
\bibitem [{\citenamefont {Ceriotti}\ \emph {et~al.}(2016)\citenamefont
  {Ceriotti}, \citenamefont {Fang}, \citenamefont {Kusalik}, \citenamefont
  {McKenzie}, \citenamefont {Michaelides}, \citenamefont {Morales},\ and\
  \citenamefont {Markland}}]{Ceriotti_Chem_Rev_2016}%
  \BibitemOpen
  \bibfield  {author} {\bibinfo {author} {\bibfnamefont {M.}~\bibnamefont
  {Ceriotti}}, \bibinfo {author} {\bibfnamefont {W.}~\bibnamefont {Fang}},
  \bibinfo {author} {\bibfnamefont {P.~G.}\ \bibnamefont {Kusalik}}, \bibinfo
  {author} {\bibfnamefont {R.~H.}\ \bibnamefont {McKenzie}}, \bibinfo {author}
  {\bibfnamefont {A.}~\bibnamefont {Michaelides}}, \bibinfo {author}
  {\bibfnamefont {M.~A.}\ \bibnamefont {Morales}},\ and\ \bibinfo {author}
  {\bibfnamefont {T.~E.}\ \bibnamefont {Markland}},\ }\href
  {https://doi.org/10.1021/acs.chemrev.5b00674} {\bibfield  {journal} {\bibinfo
   {journal} {Chem. Rev.}\ }\textbf {\bibinfo {volume} {116}},\ \bibinfo
  {pages} {7529} (\bibinfo {year} {2016})}\BibitemShut {NoStop}%
\bibitem [{\citenamefont {Pamuk}\ \emph {et~al.}(2012)\citenamefont {Pamuk},
  \citenamefont {Soler}, \citenamefont {Ram\'{\i}rez}, \citenamefont {Herrero},
  \citenamefont {Stephens}, \citenamefont {Allen},\ and\ \citenamefont
  {Fern\'andez-Serra}}]{Pamuk_PRL_2012}%
  \BibitemOpen
  \bibfield  {author} {\bibinfo {author} {\bibfnamefont {B.}~\bibnamefont
  {Pamuk}}, \bibinfo {author} {\bibfnamefont {J.~M.}\ \bibnamefont {Soler}},
  \bibinfo {author} {\bibfnamefont {R.}~\bibnamefont {Ram\'{\i}rez}}, \bibinfo
  {author} {\bibfnamefont {C.~P.}\ \bibnamefont {Herrero}}, \bibinfo {author}
  {\bibfnamefont {P.~W.}\ \bibnamefont {Stephens}}, \bibinfo {author}
  {\bibfnamefont {P.~B.}\ \bibnamefont {Allen}},\ and\ \bibinfo {author}
  {\bibfnamefont {M.-V.}\ \bibnamefont {Fern\'andez-Serra}},\ }\href
  {https://doi.org/10.1103/PhysRevLett.108.193003} {\bibfield  {journal}
  {\bibinfo  {journal} {Phys. Rev. Lett.}\ }\textbf {\bibinfo {volume} {108}},\
  \bibinfo {pages} {193003} (\bibinfo {year} {2012})}\BibitemShut {NoStop}%
\bibitem [{\citenamefont {Buxton}\ \emph {et~al.}(2019)\citenamefont {Buxton},
  \citenamefont {Quigley},\ and\ \citenamefont {Habershon}}]{Buxton_JCP_2019}%
  \BibitemOpen
  \bibfield  {author} {\bibinfo {author} {\bibfnamefont {S.~J.}\ \bibnamefont
  {Buxton}}, \bibinfo {author} {\bibfnamefont {D.}~\bibnamefont {Quigley}},\
  and\ \bibinfo {author} {\bibfnamefont {S.}~\bibnamefont {Habershon}},\ }\href
  {https://doi.org/10.1063/1.5123992} {\bibfield  {journal} {\bibinfo
  {journal} {J. Chem. Phys.}\ }\textbf {\bibinfo {volume} {151}},\ \bibinfo
  {pages} {144503} (\bibinfo {year} {2019})}\BibitemShut {NoStop}%
\bibitem [{\citenamefont {Rossi}\ \emph {et~al.}(2016)\citenamefont {Rossi},
  \citenamefont {Gasparotto},\ and\ \citenamefont {Ceriotti}}]{Rossi_PRL_2016}%
  \BibitemOpen
  \bibfield  {author} {\bibinfo {author} {\bibfnamefont {M.}~\bibnamefont
  {Rossi}}, \bibinfo {author} {\bibfnamefont {P.}~\bibnamefont {Gasparotto}},\
  and\ \bibinfo {author} {\bibfnamefont {M.}~\bibnamefont {Ceriotti}},\ }\href
  {https://doi.org/10.1103/PhysRevLett.117.115702} {\bibfield  {journal}
  {\bibinfo  {journal} {Phys. Rev. Lett.}\ }\textbf {\bibinfo {volume} {117}},\
  \bibinfo {pages} {115702} (\bibinfo {year} {2016})}\BibitemShut {NoStop}%
\bibitem [{\citenamefont {Cazorla}\ and\ \citenamefont
  {Boronat}(2017)}]{Cazorla_RevModPhys_2017}%
  \BibitemOpen
  \bibfield  {author} {\bibinfo {author} {\bibfnamefont {C.}~\bibnamefont
  {Cazorla}}\ and\ \bibinfo {author} {\bibfnamefont {J.}~\bibnamefont
  {Boronat}},\ }\href {https://doi.org/10.1103/RevModPhys.89.035003} {\bibfield
   {journal} {\bibinfo  {journal} {Rev. Mod. Phys.}\ }\textbf {\bibinfo
  {volume} {89}},\ \bibinfo {pages} {035003} (\bibinfo {year}
  {2017})}\BibitemShut {NoStop}%
\bibitem [{\citenamefont {Zacharias}\ \emph {et~al.}(2015)\citenamefont
  {Zacharias}, \citenamefont {Patrick},\ and\ \citenamefont
  {Giustino}}]{Zacharias_PRL_2015}%
  \BibitemOpen
  \bibfield  {author} {\bibinfo {author} {\bibfnamefont {M.}~\bibnamefont
  {Zacharias}}, \bibinfo {author} {\bibfnamefont {C.~E.}\ \bibnamefont
  {Patrick}},\ and\ \bibinfo {author} {\bibfnamefont {F.}~\bibnamefont
  {Giustino}},\ }\href {https://doi.org/10.1103/PhysRevLett.115.177401}
  {\bibfield  {journal} {\bibinfo  {journal} {Phys. Rev. Lett.}\ }\textbf
  {\bibinfo {volume} {115}},\ \bibinfo {pages} {177401} (\bibinfo {year}
  {2015})}\BibitemShut {NoStop}%
\bibitem [{\citenamefont {Zacharias}\ and\ \citenamefont
  {Giustino}(2016)}]{Zacharias_PRB_2016}%
  \BibitemOpen
  \bibfield  {author} {\bibinfo {author} {\bibfnamefont {M.}~\bibnamefont
  {Zacharias}}\ and\ \bibinfo {author} {\bibfnamefont {F.}~\bibnamefont
  {Giustino}},\ }\href {https://doi.org/10.1103/PhysRevB.94.075125} {\bibfield
  {journal} {\bibinfo  {journal} {Phys. Rev. B}\ }\textbf {\bibinfo {volume}
  {94}},\ \bibinfo {pages} {075125} (\bibinfo {year} {2016})}\BibitemShut
  {NoStop}%
\bibitem [{\citenamefont {Karsai}\ \emph {et~al.}(2018)\citenamefont {Karsai},
  \citenamefont {Engel}, \citenamefont {Flage-Larsen},\ and\ \citenamefont
  {Kresse}}]{Stoch_Kresse_NJP2018}%
  \BibitemOpen
  \bibfield  {author} {\bibinfo {author} {\bibfnamefont {F.}~\bibnamefont
  {Karsai}}, \bibinfo {author} {\bibfnamefont {M.}~\bibnamefont {Engel}},
  \bibinfo {author} {\bibfnamefont {E.}~\bibnamefont {Flage-Larsen}},\ and\
  \bibinfo {author} {\bibfnamefont {G.}~\bibnamefont {Kresse}},\ }\href
  {https://doi.org/10.1088/1367-2630/aaf53f} {\bibfield  {journal} {\bibinfo
  {journal} {New J. Phys.}\ }\textbf {\bibinfo {volume} {20}},\ \bibinfo
  {pages} {123008} (\bibinfo {year} {2018})}\BibitemShut {NoStop}%
\bibitem [{\citenamefont {Antonius}\ \emph {et~al.}(2015)\citenamefont
  {Antonius}, \citenamefont {Ponc\'e}, \citenamefont {Lantagne-Hurtubise},
  \citenamefont {Auclair}, \citenamefont {Gonze},\ and\ \citenamefont
  {C\^ot\'e}}]{Antonius_PRB_2015}%
  \BibitemOpen
  \bibfield  {author} {\bibinfo {author} {\bibfnamefont {G.}~\bibnamefont
  {Antonius}}, \bibinfo {author} {\bibfnamefont {S.}~\bibnamefont {Ponc\'e}},
  \bibinfo {author} {\bibfnamefont {E.}~\bibnamefont {Lantagne-Hurtubise}},
  \bibinfo {author} {\bibfnamefont {G.}~\bibnamefont {Auclair}}, \bibinfo
  {author} {\bibfnamefont {X.}~\bibnamefont {Gonze}},\ and\ \bibinfo {author}
  {\bibfnamefont {M.}~\bibnamefont {C\^ot\'e}},\ }\href
  {https://doi.org/10.1103/PhysRevB.92.085137} {\bibfield  {journal} {\bibinfo
  {journal} {Phys. Rev. B}\ }\textbf {\bibinfo {volume} {92}},\ \bibinfo
  {pages} {085137} (\bibinfo {year} {2015})}\BibitemShut {NoStop}%
\bibitem [{\citenamefont {Monserrat}\ \emph {et~al.}(2013)\citenamefont
  {Monserrat}, \citenamefont {Drummond},\ and\ \citenamefont
  {Needs}}]{VSCF_Monserrat_2013}%
  \BibitemOpen
  \bibfield  {author} {\bibinfo {author} {\bibfnamefont {B.}~\bibnamefont
  {Monserrat}}, \bibinfo {author} {\bibfnamefont {N.~D.}\ \bibnamefont
  {Drummond}},\ and\ \bibinfo {author} {\bibfnamefont {R.~J.}\ \bibnamefont
  {Needs}},\ }\href {https://doi.org/10.1103/PhysRevB.87.144302} {\bibfield
  {journal} {\bibinfo  {journal} {Phys. Rev. B}\ }\textbf {\bibinfo {volume}
  {87}},\ \bibinfo {pages} {144302} (\bibinfo {year} {2013})}\BibitemShut
  {NoStop}%
\bibitem [{\citenamefont {Giustino}\ \emph {et~al.}(2010)\citenamefont
  {Giustino}, \citenamefont {Louie},\ and\ \citenamefont
  {Cohen}}]{Giustino_PRL_2010}%
  \BibitemOpen
  \bibfield  {author} {\bibinfo {author} {\bibfnamefont {F.}~\bibnamefont
  {Giustino}}, \bibinfo {author} {\bibfnamefont {S.~G.}\ \bibnamefont
  {Louie}},\ and\ \bibinfo {author} {\bibfnamefont {M.~L.}\ \bibnamefont
  {Cohen}},\ }\href {https://doi.org/10.1103/PhysRevLett.105.265501} {\bibfield
   {journal} {\bibinfo  {journal} {Phys. Rev. Lett.}\ }\textbf {\bibinfo
  {volume} {105}},\ \bibinfo {pages} {265501} (\bibinfo {year}
  {2010})}\BibitemShut {NoStop}%
\bibitem [{\citenamefont {Cannuccia}\ and\ \citenamefont
  {Marini}(2011)}]{Canuccia_PRL_2011}%
  \BibitemOpen
  \bibfield  {author} {\bibinfo {author} {\bibfnamefont {E.}~\bibnamefont
  {Cannuccia}}\ and\ \bibinfo {author} {\bibfnamefont {A.}~\bibnamefont
  {Marini}},\ }\href {https://doi.org/10.1103/PhysRevLett.107.255501}
  {\bibfield  {journal} {\bibinfo  {journal} {Phys. Rev. Lett.}\ }\textbf
  {\bibinfo {volume} {107}},\ \bibinfo {pages} {255501} (\bibinfo {year}
  {2011})}\BibitemShut {NoStop}%
\bibitem [{\citenamefont {Kapil}\ \emph {et~al.}(2019)\citenamefont {Kapil},
  \citenamefont {Engel}, \citenamefont {Rossi},\ and\ \citenamefont
  {Ceriotti}}]{Kapil_JCTC_2019}%
  \BibitemOpen
  \bibfield  {author} {\bibinfo {author} {\bibfnamefont {V.}~\bibnamefont
  {Kapil}}, \bibinfo {author} {\bibfnamefont {E.}~\bibnamefont {Engel}},
  \bibinfo {author} {\bibfnamefont {M.}~\bibnamefont {Rossi}},\ and\ \bibinfo
  {author} {\bibfnamefont {M.}~\bibnamefont {Ceriotti}},\ }\href
  {https://doi.org/10.1021/acs.jctc.9b00596} {\bibfield  {journal} {\bibinfo
  {journal} {J. Chem. Theory Comput.}\ }\textbf {\bibinfo {volume} {15}},\
  \bibinfo {pages} {5845} (\bibinfo {year} {2019})}\BibitemShut {NoStop}%
\bibitem [{\citenamefont {Ceriotti}\ and\ \citenamefont
  {Manolopoulos}(2012)}]{PIGLET_Ceriotti_PRL_2012}%
  \BibitemOpen
  \bibfield  {author} {\bibinfo {author} {\bibfnamefont {M.}~\bibnamefont
  {Ceriotti}}\ and\ \bibinfo {author} {\bibfnamefont {D.~E.}\ \bibnamefont
  {Manolopoulos}},\ }\href {https://doi.org/10.1103/PhysRevLett.109.100604}
  {\bibfield  {journal} {\bibinfo  {journal} {Phys. Rev. Lett.}\ }\textbf
  {\bibinfo {volume} {109}},\ \bibinfo {pages} {100604} (\bibinfo {year}
  {2012})}\BibitemShut {NoStop}%
\bibitem [{\citenamefont {Ceriotti}\ \emph {et~al.}(2009)\citenamefont
  {Ceriotti}, \citenamefont {Bussi},\ and\ \citenamefont
  {Parrinello}}]{QT_Ceriotti_PRL_2009}%
  \BibitemOpen
  \bibfield  {author} {\bibinfo {author} {\bibfnamefont {M.}~\bibnamefont
  {Ceriotti}}, \bibinfo {author} {\bibfnamefont {G.}~\bibnamefont {Bussi}},\
  and\ \bibinfo {author} {\bibfnamefont {M.}~\bibnamefont {Parrinello}},\
  }\href {https://doi.org/10.1103/PhysRevLett.103.030603} {\bibfield  {journal}
  {\bibinfo  {journal} {Phys. Rev. Lett.}\ }\textbf {\bibinfo {volume} {103}},\
  \bibinfo {pages} {030603} (\bibinfo {year} {2009})}\BibitemShut {NoStop}%
\bibitem [{\citenamefont {Ram\'{\i}rez}\ \emph {et~al.}(2006)\citenamefont
  {Ram\'{\i}rez}, \citenamefont {Herrero},\ and\ \citenamefont
  {Hern\'andez}}]{PI_herrero_2006}%
  \BibitemOpen
  \bibfield  {author} {\bibinfo {author} {\bibfnamefont {R.}~\bibnamefont
  {Ram\'{\i}rez}}, \bibinfo {author} {\bibfnamefont {C.~P.}\ \bibnamefont
  {Herrero}},\ and\ \bibinfo {author} {\bibfnamefont {E.~R.}\ \bibnamefont
  {Hern\'andez}},\ }\href {https://doi.org/10.1103/PhysRevB.73.245202}
  {\bibfield  {journal} {\bibinfo  {journal} {Phys. Rev. B}\ }\textbf {\bibinfo
  {volume} {73}},\ \bibinfo {pages} {245202} (\bibinfo {year}
  {2006})}\BibitemShut {NoStop}%
\bibitem [{\citenamefont {Zacharias}\ and\ \citenamefont
  {Giustino}(2020)}]{Zacharias_PRR_2020}%
  \BibitemOpen
  \bibfield  {author} {\bibinfo {author} {\bibfnamefont {M.}~\bibnamefont
  {Zacharias}}\ and\ \bibinfo {author} {\bibfnamefont {F.}~\bibnamefont
  {Giustino}},\ }\href {https://doi.org/10.1103/PhysRevResearch.2.013357}
  {\bibfield  {journal} {\bibinfo  {journal} {Phys. Rev. Research}\ }\textbf
  {\bibinfo {volume} {2}},\ \bibinfo {pages} {013357} (\bibinfo {year}
  {2020})}\BibitemShut {NoStop}%
\bibitem [{\citenamefont {Vi\~na}\ \emph {et~al.}(1984)\citenamefont {Vi\~na},
  \citenamefont {Logothetidis},\ and\ \citenamefont {Cardona}}]{Vina_PRB_1984}%
  \BibitemOpen
  \bibfield  {author} {\bibinfo {author} {\bibfnamefont {L.}~\bibnamefont
  {Vi\~na}}, \bibinfo {author} {\bibfnamefont {S.}~\bibnamefont
  {Logothetidis}},\ and\ \bibinfo {author} {\bibfnamefont {M.}~\bibnamefont
  {Cardona}},\ }\href {https://doi.org/10.1103/PhysRevB.30.1979} {\bibfield
  {journal} {\bibinfo  {journal} {Phys. Rev. B}\ }\textbf {\bibinfo {volume}
  {30}},\ \bibinfo {pages} {1979} (\bibinfo {year} {1984})}\BibitemShut
  {NoStop}%
\bibitem [{\citenamefont {Clark}\ \emph {et~al.}(1964)\citenamefont {Clark},
  \citenamefont {Dean}, \citenamefont {Harris},\ and\ \citenamefont
  {Price}}]{dia_exp_gap_measure}%
  \BibitemOpen
  \bibfield  {author} {\bibinfo {author} {\bibfnamefont {C.~D.}\ \bibnamefont
  {Clark}}, \bibinfo {author} {\bibfnamefont {P.~J.}\ \bibnamefont {Dean}},
  \bibinfo {author} {\bibfnamefont {P.~V.}\ \bibnamefont {Harris}},\ and\
  \bibinfo {author} {\bibfnamefont {W.~C.}\ \bibnamefont {Price}},\ }\href
  {https://doi.org/10.1098/rspa.1964.0025} {\bibfield  {journal} {\bibinfo
  {journal} {Proc. Royal Soc. A Math. Phys. Sci.}\ }\textbf {\bibinfo {volume}
  {277}},\ \bibinfo {pages} {312} (\bibinfo {year} {1964})}\BibitemShut
  {NoStop}%
\bibitem [{\citenamefont {O’Donnell}\ and\ \citenamefont
  {Chen}(1991)}]{dia_exp_gap_vina_fit}%
  \BibitemOpen
  \bibfield  {author} {\bibinfo {author} {\bibfnamefont {K.~P.}\ \bibnamefont
  {O’Donnell}}\ and\ \bibinfo {author} {\bibfnamefont {X.}~\bibnamefont
  {Chen}},\ }\href {https://doi.org/10.1063/1.104723} {\bibfield  {journal}
  {\bibinfo  {journal} {Appl. Phys. Lett}\ }\textbf {\bibinfo {volume} {58}},\
  \bibinfo {pages} {2924} (\bibinfo {year} {1991})}\BibitemShut {NoStop}%
\bibitem [{\citenamefont {Yang}\ \emph {et~al.}()\citenamefont {Yang},
  \citenamefont {Govoni}, \citenamefont {Kundu},\ and\ \citenamefont
  {Galli}}]{DFPT_Han}%
  \BibitemOpen
  \bibfield  {author} {\bibinfo {author} {\bibfnamefont {H.}~\bibnamefont
  {Yang}}, \bibinfo {author} {\bibfnamefont {M.}~\bibnamefont {Govoni}},
  \bibinfo {author} {\bibfnamefont {A.}~\bibnamefont {Kundu}},\ and\ \bibinfo
  {author} {\bibfnamefont {G.}~\bibnamefont {Galli}},\ }\href@noop {} {\bibinfo
   {journal} {to be submitted}\ }\BibitemShut {NoStop}%
\bibitem [{\citenamefont {McAvoy}\ \emph {et~al.}(2018)\citenamefont {McAvoy},
  \citenamefont {Govoni},\ and\ \citenamefont {Galli}}]{DFPT_Ryan}%
  \BibitemOpen
\bibfield  {journal} {  }\bibfield  {author} {\bibinfo {author} {\bibfnamefont
  {R.~L.}\ \bibnamefont {McAvoy}}, \bibinfo {author} {\bibfnamefont
  {M.}~\bibnamefont {Govoni}},\ and\ \bibinfo {author} {\bibfnamefont
  {G.}~\bibnamefont {Galli}},\ }\href
  {https://doi.org/10.1021/acs.jctc.8b00728} {\bibfield  {journal} {\bibinfo
  {journal} {J. Chem. Theory Comput.}\ }\textbf {\bibinfo {volume} {14}},\
  \bibinfo {pages} {6269} (\bibinfo {year} {2018})}\BibitemShut {NoStop}%
\bibitem [{\citenamefont {Monserrat}(2016)}]{Thermal_lines_Monserrat}%
  \BibitemOpen
  \bibfield  {author} {\bibinfo {author} {\bibfnamefont {B.}~\bibnamefont
  {Monserrat}},\ }\href {https://doi.org/10.1103/PhysRevB.93.014302} {\bibfield
   {journal} {\bibinfo  {journal} {Phys. Rev. B}\ }\textbf {\bibinfo {volume}
  {93}},\ \bibinfo {pages} {014302} (\bibinfo {year} {2016})}\BibitemShut
  {NoStop}%
\bibitem [{\citenamefont {Piccini}\ and\ \citenamefont
  {Sauer}(2013)}]{Piccini_1}%
  \BibitemOpen
  \bibfield  {author} {\bibinfo {author} {\bibfnamefont {G.}~\bibnamefont
  {Piccini}}\ and\ \bibinfo {author} {\bibfnamefont {J.}~\bibnamefont
  {Sauer}},\ }\href {https://doi.org/10.1021/ct4005504} {\bibfield  {journal}
  {\bibinfo  {journal} {J. Chem. Theory Comput.}\ }\textbf {\bibinfo {volume}
  {9}},\ \bibinfo {pages} {5038} (\bibinfo {year} {2013})}\BibitemShut
  {NoStop}%
\bibitem [{\citenamefont {Piccini}\ and\ \citenamefont
  {Sauer}(2014)}]{Piccini_2}%
  \BibitemOpen
  \bibfield  {author} {\bibinfo {author} {\bibfnamefont {G.}~\bibnamefont
  {Piccini}}\ and\ \bibinfo {author} {\bibfnamefont {J.}~\bibnamefont
  {Sauer}},\ }\href {https://doi.org/10.1021/ct500291x} {\bibfield  {journal}
  {\bibinfo  {journal} {J. Chem. Theory Comput.}\ }\textbf {\bibinfo {volume}
  {10}},\ \bibinfo {pages} {2479} (\bibinfo {year} {2014})}\BibitemShut
  {NoStop}%
\bibitem [{\citenamefont {Errea}\ \emph {et~al.}(2014)\citenamefont {Errea},
  \citenamefont {Calandra},\ and\ \citenamefont {Mauri}}]{Mauri_PRB_2014}%
  \BibitemOpen
  \bibfield  {author} {\bibinfo {author} {\bibfnamefont {I.}~\bibnamefont
  {Errea}}, \bibinfo {author} {\bibfnamefont {M.}~\bibnamefont {Calandra}},\
  and\ \bibinfo {author} {\bibfnamefont {F.}~\bibnamefont {Mauri}},\ }\href
  {https://doi.org/10.1103/PhysRevB.89.064302} {\bibfield  {journal} {\bibinfo
  {journal} {Phys. Rev. B}\ }\textbf {\bibinfo {volume} {89}},\ \bibinfo
  {pages} {064302} (\bibinfo {year} {2014})}\BibitemShut {NoStop}%
\bibitem [{\citenamefont {Monacelli}\ and\ \citenamefont
  {Mauri}(2020)}]{Monacelli_preprint_2020}%
  \BibitemOpen
  \bibfield  {author} {\bibinfo {author} {\bibfnamefont {L.}~\bibnamefont
  {Monacelli}}\ and\ \bibinfo {author} {\bibfnamefont {F.}~\bibnamefont
  {Mauri}},\ }\href@noop {} {\bibinfo {title} {Time-dependent self consistent
  harmonic approximation: Anharmonic nuclear quantum dynamics and time
  correlation functions}} (\bibinfo {year} {2020}),\ \Eprint
  {https://arxiv.org/abs/2011.14986} {arXiv:2011.14986} \BibitemShut {NoStop}%
\bibitem [{\citenamefont {García-Risueño}\ \emph {et~al.}(2019)\citenamefont
  {García-Risueño}, \citenamefont {Han},\ and\ \citenamefont
  {Bester}}]{FPH_Bester_2019}%
  \BibitemOpen
  \bibfield  {author} {\bibinfo {author} {\bibfnamefont {P.}~\bibnamefont
  {García-Risueño}}, \bibinfo {author} {\bibfnamefont {P.}~\bibnamefont
  {Han}},\ and\ \bibinfo {author} {\bibfnamefont {G.}~\bibnamefont {Bester}},\
  }\href@noop {} {\bibinfo {title} {Frozen-phonon method for state anticrossing
  situations and its application to zero-point motion effects in diamondoids}}
  (\bibinfo {year} {2019}),\ \Eprint {https://arxiv.org/abs/1904.05385}
  {arXiv:1904.05385} \BibitemShut {NoStop}%
\bibitem [{\citenamefont {Patrick}\ and\ \citenamefont
  {Giustino}(2013)}]{Giustino_Nat_Commun_2013}%
  \BibitemOpen
  \bibfield  {author} {\bibinfo {author} {\bibfnamefont {C.~E.}\ \bibnamefont
  {Patrick}}\ and\ \bibinfo {author} {\bibfnamefont {F.}~\bibnamefont
  {Giustino}},\ }\href {https://doi.org/10.1038/ncomms3006} {\bibfield
  {journal} {\bibinfo  {journal} {Nat. Commun}\ }\textbf {\bibinfo {volume}
  {4}},\ \bibinfo {pages} {2006} (\bibinfo {year} {2013})}\BibitemShut
  {NoStop}%
\bibitem [{\citenamefont {Gali}\ \emph {et~al.}(2016)\citenamefont {Gali},
  \citenamefont {Demj{\'a}n}, \citenamefont {V{\"o}r{\"o}s}, \citenamefont
  {Thiering}, \citenamefont {Cannuccia},\ and\ \citenamefont
  {Marini}}]{Gali_Nat_Commun_2016}%
  \BibitemOpen
  \bibfield  {author} {\bibinfo {author} {\bibfnamefont {A.}~\bibnamefont
  {Gali}}, \bibinfo {author} {\bibfnamefont {T.}~\bibnamefont {Demj{\'a}n}},
  \bibinfo {author} {\bibfnamefont {M.}~\bibnamefont {V{\"o}r{\"o}s}}, \bibinfo
  {author} {\bibfnamefont {G.}~\bibnamefont {Thiering}}, \bibinfo {author}
  {\bibfnamefont {E.}~\bibnamefont {Cannuccia}},\ and\ \bibinfo {author}
  {\bibfnamefont {A.}~\bibnamefont {Marini}},\ }\href
  {https://doi.org/10.1038/ncomms11327} {\bibfield  {journal} {\bibinfo
  {journal} {Nat. Commun}\ }\textbf {\bibinfo {volume} {7}},\ \bibinfo {pages}
  {11327} (\bibinfo {year} {2016})}\BibitemShut {NoStop}%
\bibitem [{\citenamefont {Govoni}\ and\ \citenamefont
  {Galli}(2015)}]{Govoni2015}%
  \BibitemOpen
  \bibfield  {author} {\bibinfo {author} {\bibfnamefont {M.}~\bibnamefont
  {Govoni}}\ and\ \bibinfo {author} {\bibfnamefont {G.}~\bibnamefont {Galli}},\
  }\href {https://doi.org/10.1021/ct500958p} {\bibfield  {journal} {\bibinfo
  {journal} {J. Chem. Theory Comput.}\ }\textbf {\bibinfo {volume} {11}},\
  \bibinfo {pages} {2680} (\bibinfo {year} {2015})}\BibitemShut {NoStop}%
\bibitem [{\citenamefont {Risplendi}\ \emph {et~al.}(2014)\citenamefont
  {Risplendi}, \citenamefont {Bernardi}, \citenamefont {Cicero},\ and\
  \citenamefont {Grossman}}]{aC_Risplendi_APL_2014}%
  \BibitemOpen
  \bibfield  {author} {\bibinfo {author} {\bibfnamefont {F.}~\bibnamefont
  {Risplendi}}, \bibinfo {author} {\bibfnamefont {M.}~\bibnamefont {Bernardi}},
  \bibinfo {author} {\bibfnamefont {G.}~\bibnamefont {Cicero}},\ and\ \bibinfo
  {author} {\bibfnamefont {J.~C.}\ \bibnamefont {Grossman}},\ }\href
  {https://doi.org/10.1063/1.4891498} {\bibfield  {journal} {\bibinfo
  {journal} {Appl. Phys. Lett}\ }\textbf {\bibinfo {volume} {105}},\ \bibinfo
  {pages} {043903} (\bibinfo {year} {2014})}\BibitemShut {NoStop}%
\bibitem [{\citenamefont {Iwasawa}\ \emph {et~al.}(2018)\citenamefont
  {Iwasawa}, \citenamefont {Takita}, \citenamefont {Goto}, \citenamefont
  {Mansuer}, \citenamefont {Miyashita}, \citenamefont {Schwier}, \citenamefont
  {Ino}, \citenamefont {Shimada},\ and\ \citenamefont
  {Aiura}}]{ARPES_Iwasawa2018}%
  \BibitemOpen
  \bibfield  {author} {\bibinfo {author} {\bibfnamefont {H.}~\bibnamefont
  {Iwasawa}}, \bibinfo {author} {\bibfnamefont {H.}~\bibnamefont {Takita}},
  \bibinfo {author} {\bibfnamefont {K.}~\bibnamefont {Goto}}, \bibinfo {author}
  {\bibfnamefont {W.}~\bibnamefont {Mansuer}}, \bibinfo {author} {\bibfnamefont
  {T.}~\bibnamefont {Miyashita}}, \bibinfo {author} {\bibfnamefont {E.~F.}\
  \bibnamefont {Schwier}}, \bibinfo {author} {\bibfnamefont {A.}~\bibnamefont
  {Ino}}, \bibinfo {author} {\bibfnamefont {K.}~\bibnamefont {Shimada}},\ and\
  \bibinfo {author} {\bibfnamefont {Y.}~\bibnamefont {Aiura}},\ }\href
  {https://doi.org/10.1038/s41598-018-34894-7} {\bibfield  {journal} {\bibinfo
  {journal} {Sci. Rep.}\ }\textbf {\bibinfo {volume} {8}},\ \bibinfo {pages}
  {17431} (\bibinfo {year} {2018})}\BibitemShut {NoStop}%
\end{thebibliography}%


\providecommand{\noopsort}[1]{}\providecommand{\singleletter}[1]{#1}%
\begin{thebibliography}{24}%
\makeatletter
\providecommand \@ifxundefined [1]{%
 \@ifx{#1\undefined}
}%
\providecommand \@ifnum [1]{%
 \ifnum #1\expandafter \@firstoftwo
 \else \expandafter \@secondoftwo
 \fi
}%
\providecommand \@ifx [1]{%
 \ifx #1\expandafter \@firstoftwo
 \else \expandafter \@secondoftwo
 \fi
}%
\providecommand \natexlab [1]{#1}%
\providecommand \enquote  [1]{``#1''}%
\providecommand \bibnamefont  [1]{#1}%
\providecommand \bibfnamefont [1]{#1}%
\providecommand \citenamefont [1]{#1}%
\providecommand \href@noop [0]{\@secondoftwo}%
\providecommand \href [0]{\begingroup \@sanitize@url \@href}%
\providecommand \@href[1]{\@@startlink{#1}\@@href}%
\providecommand \@@href[1]{\endgroup#1\@@endlink}%
\providecommand \@sanitize@url [0]{\catcode `\\12\catcode `\$12\catcode
  `\&12\catcode `\#12\catcode `\^12\catcode `\_12\catcode `\%12\relax}%
\providecommand \@@startlink[1]{}%
\providecommand \@@endlink[0]{}%
\providecommand \url  [0]{\begingroup\@sanitize@url \@url }%
\providecommand \@url [1]{\endgroup\@href {#1}{\urlprefix }}%
\providecommand \urlprefix  [0]{URL }%
\providecommand \Eprint [0]{\href }%
\providecommand \doibase [0]{https://doi.org/}%
\providecommand \selectlanguage [0]{\@gobble}%
\providecommand \bibinfo  [0]{\@secondoftwo}%
\providecommand \bibfield  [0]{\@secondoftwo}%
\providecommand \translation [1]{[#1]}%
\providecommand \BibitemOpen [0]{}%
\providecommand \bibitemStop [0]{}%
\providecommand \bibitemNoStop [0]{.\EOS\space}%
\providecommand \EOS [0]{\spacefactor3000\relax}%
\providecommand \BibitemShut  [1]{\csname bibitem#1\endcsname}%
\let\auto@bib@innerbib\@empty
%</preamble>
\bibitem [{\citenamefont {Kapil}\ \emph {et~al.}(2019)\citenamefont {Kapil},
  \citenamefont {Rossi}, \citenamefont {Marsalek}, \citenamefont {Petraglia},
  \citenamefont {Litman}, \citenamefont {Spura}, \citenamefont {Cheng},
  \citenamefont {Cuzzocrea}, \citenamefont {Meißner}, \citenamefont {Wilkins},
  \citenamefont {Helfrecht}, \citenamefont {Juda}, \citenamefont {Bienvenue},
  \citenamefont {Fang}, \citenamefont {Kessler}, \citenamefont {Poltavsky},
  \citenamefont {Vandenbrande}, \citenamefont {Wieme}, \citenamefont
  {Corminboeuf}, \citenamefont {Kühne}, \citenamefont {Manolopoulos},
  \citenamefont {Markland}, \citenamefont {Richardson}, \citenamefont
  {Tkatchenko}, \citenamefont {Tribello}, \citenamefont {{Van Speybroeck}},\
  and\ \citenamefont {Ceriotti}}]{ipi_Kapil_2018}%
  \BibitemOpen
  \bibfield  {author} {\bibinfo {author} {\bibfnamefont {V.}~\bibnamefont
  {Kapil}}, \bibinfo {author} {\bibfnamefont {M.}~\bibnamefont {Rossi}},
  \bibinfo {author} {\bibfnamefont {O.}~\bibnamefont {Marsalek}}, \bibinfo
  {author} {\bibfnamefont {R.}~\bibnamefont {Petraglia}}, \bibinfo {author}
  {\bibfnamefont {Y.}~\bibnamefont {Litman}}, \bibinfo {author} {\bibfnamefont
  {T.}~\bibnamefont {Spura}}, \bibinfo {author} {\bibfnamefont
  {B.}~\bibnamefont {Cheng}}, \bibinfo {author} {\bibfnamefont
  {A.}~\bibnamefont {Cuzzocrea}}, \bibinfo {author} {\bibfnamefont {R.~H.}\
  \bibnamefont {Meißner}}, \bibinfo {author} {\bibfnamefont {D.~M.}\
  \bibnamefont {Wilkins}}, \bibinfo {author} {\bibfnamefont {B.~A.}\
  \bibnamefont {Helfrecht}}, \bibinfo {author} {\bibfnamefont {P.}~\bibnamefont
  {Juda}}, \bibinfo {author} {\bibfnamefont {S.~P.}\ \bibnamefont {Bienvenue}},
  \bibinfo {author} {\bibfnamefont {W.}~\bibnamefont {Fang}}, \bibinfo {author}
  {\bibfnamefont {J.}~\bibnamefont {Kessler}}, \bibinfo {author} {\bibfnamefont
  {I.}~\bibnamefont {Poltavsky}}, \bibinfo {author} {\bibfnamefont
  {S.}~\bibnamefont {Vandenbrande}}, \bibinfo {author} {\bibfnamefont
  {J.}~\bibnamefont {Wieme}}, \bibinfo {author} {\bibfnamefont
  {C.}~\bibnamefont {Corminboeuf}}, \bibinfo {author} {\bibfnamefont {T.~D.}\
  \bibnamefont {Kühne}}, \bibinfo {author} {\bibfnamefont {D.~E.}\
  \bibnamefont {Manolopoulos}}, \bibinfo {author} {\bibfnamefont {T.~E.}\
  \bibnamefont {Markland}}, \bibinfo {author} {\bibfnamefont {J.~O.}\
  \bibnamefont {Richardson}}, \bibinfo {author} {\bibfnamefont
  {A.}~\bibnamefont {Tkatchenko}}, \bibinfo {author} {\bibfnamefont {G.~A.}\
  \bibnamefont {Tribello}}, \bibinfo {author} {\bibfnamefont {V.}~\bibnamefont
  {{Van Speybroeck}}},\ and\ \bibinfo {author} {\bibfnamefont {M.}~\bibnamefont
  {Ceriotti}},\ }\href
  {https://doi.org/https://doi.org/10.1016/j.cpc.2018.09.020} {\bibfield
  {journal} {\bibinfo  {journal} {Comput. Phys. Commun}\ }\textbf {\bibinfo
  {volume} {236}},\ \bibinfo {pages} {214 } (\bibinfo {year}
  {2019})}\BibitemShut {NoStop}%
\bibitem [{\citenamefont {{Gygi}}(2008)}]{Qbox_Gygi_2008}%
  \BibitemOpen
  \bibfield  {author} {\bibinfo {author} {\bibfnamefont {F.}~\bibnamefont
  {{Gygi}}},\ }\href {https://doi.org/10.1147/rd.521.0137} {\bibfield
  {journal} {\bibinfo  {journal} {IBM J. Res. Dev}\ }\textbf {\bibinfo {volume}
  {52}},\ \bibinfo {pages} {137} (\bibinfo {year} {2008})}\BibitemShut
  {NoStop}%
\bibitem [{\citenamefont {Ceriotti}\ \emph {et~al.}(2009)\citenamefont
  {Ceriotti}, \citenamefont {Bussi},\ and\ \citenamefont
  {Parrinello}}]{QT_Ceriotti_PRL_2009}%
  \BibitemOpen
  \bibfield  {author} {\bibinfo {author} {\bibfnamefont {M.}~\bibnamefont
  {Ceriotti}}, \bibinfo {author} {\bibfnamefont {G.}~\bibnamefont {Bussi}},\
  and\ \bibinfo {author} {\bibfnamefont {M.}~\bibnamefont {Parrinello}},\
  }\href {https://doi.org/10.1103/PhysRevLett.103.030603} {\bibfield  {journal}
  {\bibinfo  {journal} {Phys. Rev. Lett.}\ }\textbf {\bibinfo {volume} {103}},\
  \bibinfo {pages} {030603} (\bibinfo {year} {2009})}\BibitemShut {NoStop}%
\bibitem [{\citenamefont {Ceriotti}\ \emph {et~al.}(2010)\citenamefont
  {Ceriotti}, \citenamefont {Bussi},\ and\ \citenamefont
  {Parrinello}}]{QT_Ceriotti_JCTC_2010}%
  \BibitemOpen
  \bibfield  {author} {\bibinfo {author} {\bibfnamefont {M.}~\bibnamefont
  {Ceriotti}}, \bibinfo {author} {\bibfnamefont {G.}~\bibnamefont {Bussi}},\
  and\ \bibinfo {author} {\bibfnamefont {M.}~\bibnamefont {Parrinello}},\
  }\href {https://doi.org/10.1021/ct900563s} {\bibfield  {journal} {\bibinfo
  {journal} {J. Chem. Theory Comput.}\ }\textbf {\bibinfo {volume} {6}},\
  \bibinfo {pages} {1170} (\bibinfo {year} {2010})}\BibitemShut {NoStop}%
\bibitem [{\citenamefont {Ceriotti}\ and\ \citenamefont
  {Manolopoulos}(2012)}]{PIGLET_Ceriotti_PRL_2012}%
  \BibitemOpen
  \bibfield  {author} {\bibinfo {author} {\bibfnamefont {M.}~\bibnamefont
  {Ceriotti}}\ and\ \bibinfo {author} {\bibfnamefont {D.~E.}\ \bibnamefont
  {Manolopoulos}},\ }\href {https://doi.org/10.1103/PhysRevLett.109.100604}
  {\bibfield  {journal} {\bibinfo  {journal} {Phys. Rev. Lett.}\ }\textbf
  {\bibinfo {volume} {109}},\ \bibinfo {pages} {100604} (\bibinfo {year}
  {2012})}\BibitemShut {NoStop}%
\bibitem [{\citenamefont {Berne}\ and\ \citenamefont
  {Thirumalai}(1986)}]{PI_Berne_Rev_1986}%
  \BibitemOpen
  \bibfield  {author} {\bibinfo {author} {\bibfnamefont {B.~J.}\ \bibnamefont
  {Berne}}\ and\ \bibinfo {author} {\bibfnamefont {D.}~\bibnamefont
  {Thirumalai}},\ }\href {https://doi.org/10.1146/annurev.pc.37.100186.002153}
  {\bibfield  {journal} {\bibinfo  {journal} {Annu. Rev. Phys. Chem.}\ }\textbf
  {\bibinfo {volume} {37}},\ \bibinfo {pages} {401} (\bibinfo {year}
  {1986})}\BibitemShut {NoStop}%
\bibitem [{\citenamefont {Marx}\ and\ \citenamefont
  {Parrinello}(1996)}]{PI_Marx_Rev_1996}%
  \BibitemOpen
  \bibfield  {author} {\bibinfo {author} {\bibfnamefont {D.}~\bibnamefont
  {Marx}}\ and\ \bibinfo {author} {\bibfnamefont {M.}~\bibnamefont
  {Parrinello}},\ }\href {https://doi.org/10.1063/1.471221} {\bibfield
  {journal} {\bibinfo  {journal} {J. Chem. Phys.}\ }\textbf {\bibinfo {volume}
  {104}},\ \bibinfo {pages} {4077} (\bibinfo {year} {1996})}\BibitemShut
  {NoStop}%
\bibitem [{\citenamefont {Herrero}\ and\ \citenamefont
  {Ram{\'{\i}}rez}(2014)}]{PI_hererro_rev_2014}%
  \BibitemOpen
  \bibfield  {author} {\bibinfo {author} {\bibfnamefont {C.~P.}\ \bibnamefont
  {Herrero}}\ and\ \bibinfo {author} {\bibfnamefont {R.}~\bibnamefont
  {Ram{\'{\i}}rez}},\ }\href {https://doi.org/10.1088/0953-8984/26/23/233201}
  {\bibfield  {journal} {\bibinfo  {journal} {J. Phys. Condens. Matter}\
  }\textbf {\bibinfo {volume} {26}},\ \bibinfo {pages} {233201} (\bibinfo
  {year} {2014})}\BibitemShut {NoStop}%
\bibitem [{\citenamefont {Perdew}\ \emph {et~al.}(1996)\citenamefont {Perdew},
  \citenamefont {Burke},\ and\ \citenamefont
  {Ernzerhof}}]{PBE_Perdew_PRL_1996_1}%
  \BibitemOpen
  \bibfield  {author} {\bibinfo {author} {\bibfnamefont {J.~P.}\ \bibnamefont
  {Perdew}}, \bibinfo {author} {\bibfnamefont {K.}~\bibnamefont {Burke}},\ and\
  \bibinfo {author} {\bibfnamefont {M.}~\bibnamefont {Ernzerhof}},\ }\href
  {https://doi.org/10.1103/PhysRevLett.77.3865} {\bibfield  {journal} {\bibinfo
   {journal} {Phys. Rev. Lett.}\ }\textbf {\bibinfo {volume} {77}},\ \bibinfo
  {pages} {3865} (\bibinfo {year} {1996})}\BibitemShut {NoStop}%
\bibitem [{\citenamefont {Perdew}\ \emph {et~al.}(1997)\citenamefont {Perdew},
  \citenamefont {Burke},\ and\ \citenamefont
  {Ernzerhof}}]{PBE_Perdew_PRL_1996_2}%
  \BibitemOpen
  \bibfield  {author} {\bibinfo {author} {\bibfnamefont {J.~P.}\ \bibnamefont
  {Perdew}}, \bibinfo {author} {\bibfnamefont {K.}~\bibnamefont {Burke}},\ and\
  \bibinfo {author} {\bibfnamefont {M.}~\bibnamefont {Ernzerhof}},\ }\href
  {https://doi.org/10.1103/PhysRevLett.78.1396} {\bibfield  {journal} {\bibinfo
   {journal} {Phys. Rev. Lett.}\ }\textbf {\bibinfo {volume} {78}},\ \bibinfo
  {pages} {1396} (\bibinfo {year} {1997})}\BibitemShut {NoStop}%
\bibitem [{\citenamefont {Schlipf}\ and\ \citenamefont
  {Gygi}(2015)}]{ONCV_2015}%
  \BibitemOpen
  \bibfield  {author} {\bibinfo {author} {\bibfnamefont {M.}~\bibnamefont
  {Schlipf}}\ and\ \bibinfo {author} {\bibfnamefont {F.}~\bibnamefont {Gygi}},\
  }\href {https://doi.org/https://doi.org/10.1016/j.cpc.2015.05.011} {\bibfield
   {journal} {\bibinfo  {journal} {Comput. Phys. Commun}\ }\textbf {\bibinfo
  {volume} {196}},\ \bibinfo {pages} {36 } (\bibinfo {year}
  {2015})}\BibitemShut {NoStop}%
\bibitem [{\citenamefont {Bussi}\ \emph {et~al.}(2007)\citenamefont {Bussi},
  \citenamefont {Donadio},\ and\ \citenamefont {Parrinello}}]{BDP_JCP_2007}%
  \BibitemOpen
  \bibfield  {author} {\bibinfo {author} {\bibfnamefont {G.}~\bibnamefont
  {Bussi}}, \bibinfo {author} {\bibfnamefont {D.}~\bibnamefont {Donadio}},\
  and\ \bibinfo {author} {\bibfnamefont {M.}~\bibnamefont {Parrinello}},\
  }\href {https://doi.org/10.1063/1.2408420} {\bibfield  {journal} {\bibinfo
  {journal} {J. Chem. Phys.}\ }\textbf {\bibinfo {volume} {126}},\ \bibinfo
  {pages} {014101} (\bibinfo {year} {2007})}\BibitemShut {NoStop}%
\bibitem [{\citenamefont {Vi\~na}\ \emph {et~al.}(1984)\citenamefont {Vi\~na},
  \citenamefont {Logothetidis},\ and\ \citenamefont {Cardona}}]{Vina_PRB_1984}%
  \BibitemOpen
  \bibfield  {author} {\bibinfo {author} {\bibfnamefont {L.}~\bibnamefont
  {Vi\~na}}, \bibinfo {author} {\bibfnamefont {S.}~\bibnamefont
  {Logothetidis}},\ and\ \bibinfo {author} {\bibfnamefont {M.}~\bibnamefont
  {Cardona}},\ }\href {https://doi.org/10.1103/PhysRevB.30.1979} {\bibfield
  {journal} {\bibinfo  {journal} {Phys. Rev. B}\ }\textbf {\bibinfo {volume}
  {30}},\ \bibinfo {pages} {1979} (\bibinfo {year} {1984})}\BibitemShut
  {NoStop}%
\bibitem [{\citenamefont {Han}\ and\ \citenamefont
  {Bester}(2013)}]{FPH_Bester_2013}%
  \BibitemOpen
  \bibfield  {author} {\bibinfo {author} {\bibfnamefont {P.}~\bibnamefont
  {Han}}\ and\ \bibinfo {author} {\bibfnamefont {G.}~\bibnamefont {Bester}},\
  }\href {https://doi.org/10.1103/PhysRevB.88.165311} {\bibfield  {journal}
  {\bibinfo  {journal} {Phys. Rev. B}\ }\textbf {\bibinfo {volume} {88}},\
  \bibinfo {pages} {165311} (\bibinfo {year} {2013})}\BibitemShut {NoStop}%
\bibitem [{\citenamefont {Antonius}\ \emph {et~al.}(2015)\citenamefont
  {Antonius}, \citenamefont {Ponc\'e}, \citenamefont {Lantagne-Hurtubise},
  \citenamefont {Auclair}, \citenamefont {Gonze},\ and\ \citenamefont
  {C\^ot\'e}}]{Antonius_PRB_2015}%
  \BibitemOpen
  \bibfield  {author} {\bibinfo {author} {\bibfnamefont {G.}~\bibnamefont
  {Antonius}}, \bibinfo {author} {\bibfnamefont {S.}~\bibnamefont {Ponc\'e}},
  \bibinfo {author} {\bibfnamefont {E.}~\bibnamefont {Lantagne-Hurtubise}},
  \bibinfo {author} {\bibfnamefont {G.}~\bibnamefont {Auclair}}, \bibinfo
  {author} {\bibfnamefont {X.}~\bibnamefont {Gonze}},\ and\ \bibinfo {author}
  {\bibfnamefont {M.}~\bibnamefont {C\^ot\'e}},\ }\href
  {https://doi.org/10.1103/PhysRevB.92.085137} {\bibfield  {journal} {\bibinfo
  {journal} {Phys. Rev. B}\ }\textbf {\bibinfo {volume} {92}},\ \bibinfo
  {pages} {085137} (\bibinfo {year} {2015})}\BibitemShut {NoStop}%
\bibitem [{\citenamefont {Han}\ and\ \citenamefont
  {Bester}(2016)}]{FPH_Bester_2016}%
  \BibitemOpen
  \bibfield  {author} {\bibinfo {author} {\bibfnamefont {P.}~\bibnamefont
  {Han}}\ and\ \bibinfo {author} {\bibfnamefont {G.}~\bibnamefont {Bester}},\
  }\href {https://doi.org/10.1088/1367-2630/18/11/113052} {\bibfield  {journal}
  {\bibinfo  {journal} {New Journal of Physics}\ }\textbf {\bibinfo {volume}
  {18}},\ \bibinfo {pages} {113052} (\bibinfo {year} {2016})}\BibitemShut
  {NoStop}%
\bibitem [{\citenamefont {Monserrat}(2018)}]{Monserrat_Rev_2018}%
  \BibitemOpen
  \bibfield  {author} {\bibinfo {author} {\bibfnamefont {B.}~\bibnamefont
  {Monserrat}},\ }\href {https://doi.org/10.1088/1361-648x/aaa737} {\bibfield
  {journal} {\bibinfo  {journal} {J. Phys. Condens. Matter}\ }\textbf {\bibinfo
  {volume} {30}},\ \bibinfo {pages} {083001} (\bibinfo {year}
  {2018})}\BibitemShut {NoStop}%
\bibitem [{\citenamefont {García-Risueño}\ \emph {et~al.}(2019)\citenamefont
  {García-Risueño}, \citenamefont {Han},\ and\ \citenamefont
  {Bester}}]{FPH_Bester_2019}%
  \BibitemOpen
  \bibfield  {author} {\bibinfo {author} {\bibfnamefont {P.}~\bibnamefont
  {García-Risueño}}, \bibinfo {author} {\bibfnamefont {P.}~\bibnamefont
  {Han}},\ and\ \bibinfo {author} {\bibfnamefont {G.}~\bibnamefont {Bester}},\
  }\href@noop {} {\bibinfo {title} {Frozen-phonon method for state anticrossing
  situations and its application to zero-point motion effects in diamondoids}}
  (\bibinfo {year} {2019}),\ \Eprint {https://arxiv.org/abs/1904.05385}
  {arXiv:1904.05385} \BibitemShut {NoStop}%
\bibitem [{\citenamefont {D'Agostino}\ and\ \citenamefont
  {Belanger}(1990)}]{skewtest2}%
  \BibitemOpen
  \bibfield  {author} {\bibinfo {author} {\bibfnamefont {R.~B.}\ \bibnamefont
  {D'Agostino}}\ and\ \bibinfo {author} {\bibfnamefont {A.}~\bibnamefont
  {Belanger}},\ }\href {http://www.jstor.org/stable/2684359} {\bibfield
  {journal} {\bibinfo  {journal} {Am. Stat.}\ }\textbf {\bibinfo {volume}
  {44}},\ \bibinfo {pages} {316} (\bibinfo {year} {1990})}\BibitemShut
  {NoStop}%
\bibitem [{\citenamefont {Zwillinger}\ and\ \citenamefont
  {Kokoska}(2000)}]{CRC_stat_book}%
  \BibitemOpen
  \bibfield  {author} {\bibinfo {author} {\bibfnamefont {D.}~\bibnamefont
  {Zwillinger}}\ and\ \bibinfo {author} {\bibfnamefont {S.}~\bibnamefont
  {Kokoska}},\ }\href@noop {} {\emph {\bibinfo {title} {CRC standard
  probability and statistics tables and formulae}}}\ (\bibinfo  {publisher}
  {Chapman {\&} Hall/CRC},\ \bibinfo {address} {Boca Raton},\ \bibinfo {year}
  {2000})\BibitemShut {NoStop}%
\bibitem [{\citenamefont {Du}()}]{order_code}%
  \BibitemOpen
  \bibfield  {author} {\bibinfo {author} {\bibfnamefont {P.}~\bibnamefont
  {Du}},\ }\href@noop {} {\bibinfo {title} {Order: a tool to characterize the
  local structure of liquid water by geometric order parameters}},\ \bibinfo
  {howpublished}
  {\url{https://order.readthedocs.io/en/latest/index.html}}\BibitemShut
  {NoStop}%
\bibitem [{\citenamefont {Chau}\ and\ \citenamefont
  {Hardwick}(1998)}]{oto_ref_mol_phys_1998}%
  \BibitemOpen
  \bibfield  {author} {\bibinfo {author} {\bibfnamefont {P.~L.}\ \bibnamefont
  {Chau}}\ and\ \bibinfo {author} {\bibfnamefont {A.~J.}\ \bibnamefont
  {Hardwick}},\ }\href {https://doi.org/10.1080/002689798169195} {\bibfield
  {journal} {\bibinfo  {journal} {Mol. Phys.}\ }\textbf {\bibinfo {volume}
  {93}},\ \bibinfo {pages} {511} (\bibinfo {year} {1998})}\BibitemShut
  {NoStop}%
\bibitem [{\citenamefont {Errington}\ and\ \citenamefont
  {Debenedetti}(2001)}]{oto_ref_Errington_Nat_2001}%
  \BibitemOpen
  \bibfield  {author} {\bibinfo {author} {\bibfnamefont {J.~R.}\ \bibnamefont
  {Errington}}\ and\ \bibinfo {author} {\bibfnamefont {P.~G.}\ \bibnamefont
  {Debenedetti}},\ }\href {https://doi.org/10.1038/35053024} {\bibfield
  {journal} {\bibinfo  {journal} {Nature}\ }\textbf {\bibinfo {volume} {409}},\
  \bibinfo {pages} {318} (\bibinfo {year} {2001})}\BibitemShut {NoStop}%
\bibitem [{\citenamefont {Mounet}\ and\ \citenamefont
  {Marzari}(2005)}]{QHA_Marzari_2005}%
  \BibitemOpen
  \bibfield  {author} {\bibinfo {author} {\bibfnamefont {N.}~\bibnamefont
  {Mounet}}\ and\ \bibinfo {author} {\bibfnamefont {N.}~\bibnamefont
  {Marzari}},\ }\href {https://doi.org/10.1103/PhysRevB.71.205214} {\bibfield
  {journal} {\bibinfo  {journal} {Phys. Rev. B}\ }\textbf {\bibinfo {volume}
  {71}},\ \bibinfo {pages} {205214} (\bibinfo {year} {2005})}\BibitemShut
  {NoStop}%
\end{thebibliography}%
%TC:endignore

% Don't count these!
%TC:ignore
%\newpage
%\section{Words Count details}
%\noindent
%\textbf{Figures details}\\
%Figure-1: AR = 4.00, 2-column, Word count = \(300/(0.5*4)+40 = 190\)\\
%Figure-2: AR = 2.00, 1-column, Word count = \(150/2+20 = 95\)\\
%Figure-3: AR = 1.67, 1-column, Word count = \(150/1.67+20 = 110\)\\
%Figure-4: AR = 1.67, 1-column, Word count = \(150/1.67+20 = 110\)\\
%-----------------------------------------------------------------------\\
%Total words for figures = 505\\
%-----------------------------------------------------------------------\\
%\textbf{Texts details}\\
%\detailtexcount{}
%TC:endignore

\end{document}

%\preprint{APS/123-QED}

\title{SUPPLEMENTAL MATERIAL\\ Quantum vibronic effects on the electronic properties of solid and molecular carbon}% Force line breaks with \\

\author{Arpan Kundu}
\email[Corresponding author. Email: ]{arpank@uchicago.edu}
\affiliation{Pritzker School of Molecular Engineering, University of Chicago, Chicago, Illinois 60637, United States}

\author{Marco Govoni}
\affiliation{ Materials Science Division and Center for Molecular Engineering, Argonne National Laboratory, Lemont, Illinois 60439, United States.}
\affiliation{Pritzker School of Molecular Engineering, University of Chicago, Chicago, Illinois 60637, United States}

\author {Han Yang}
\affiliation{Department of Chemistry, University of Chicago, Chicago, Illinois 60637, United States}

\author{Michele Ceriotti}
\affiliation{Laboratory of Computational Science and Modeling, IMX, Ecole Polytechnique Federale de Lausanne, 1015 Lausanne, Switzerland}

\author{Francois Gygi}
\affiliation{Department of Computer Science, University of California Davis, Davis, California 95616, United States}

\author{Giulia Galli}
\email[Corresponding author. Email: ]{gagalli@uchicago.edu}
 \affiliation{Pritzker School of Molecular Engineering, University of Chicago, Chicago, Illinois 60637, United States}
 \affiliation{ Materials Science Division and Center for Molecular Engineering, Argonne National Laboratory, Lemont, Illinois 60439, United States.}
\affiliation{Department of Chemistry, University of Chicago, Chicago, Illinois 60637, United States}


\maketitle

\tableofcontents

\newpage
%Parts moved from maintext

%Compared to a classical MD, a remarkable improvement was achieved for reproducing NQE in structural properties; however, a considerable discrepancy was still observed for systems with strong anharmonicity due to zero-point energy leakage \cite{QT_Ceriotti_PRL_2009}. Later, it was found that in a PI method if GLE is attached to each normal mode of the beads while a classical thermostat is attached to the centroid, the resulting PI method known as ``PIGLET" can correct the zero-point energy leakage problem and can drastically reduce the bead requirements \cite{PIGLET_Ceriotti_PRL_2012}.

\section{iPI---Qbox coupling}
\begin{figure}[H]
    \centering
    \includegraphics[scale=0.5]{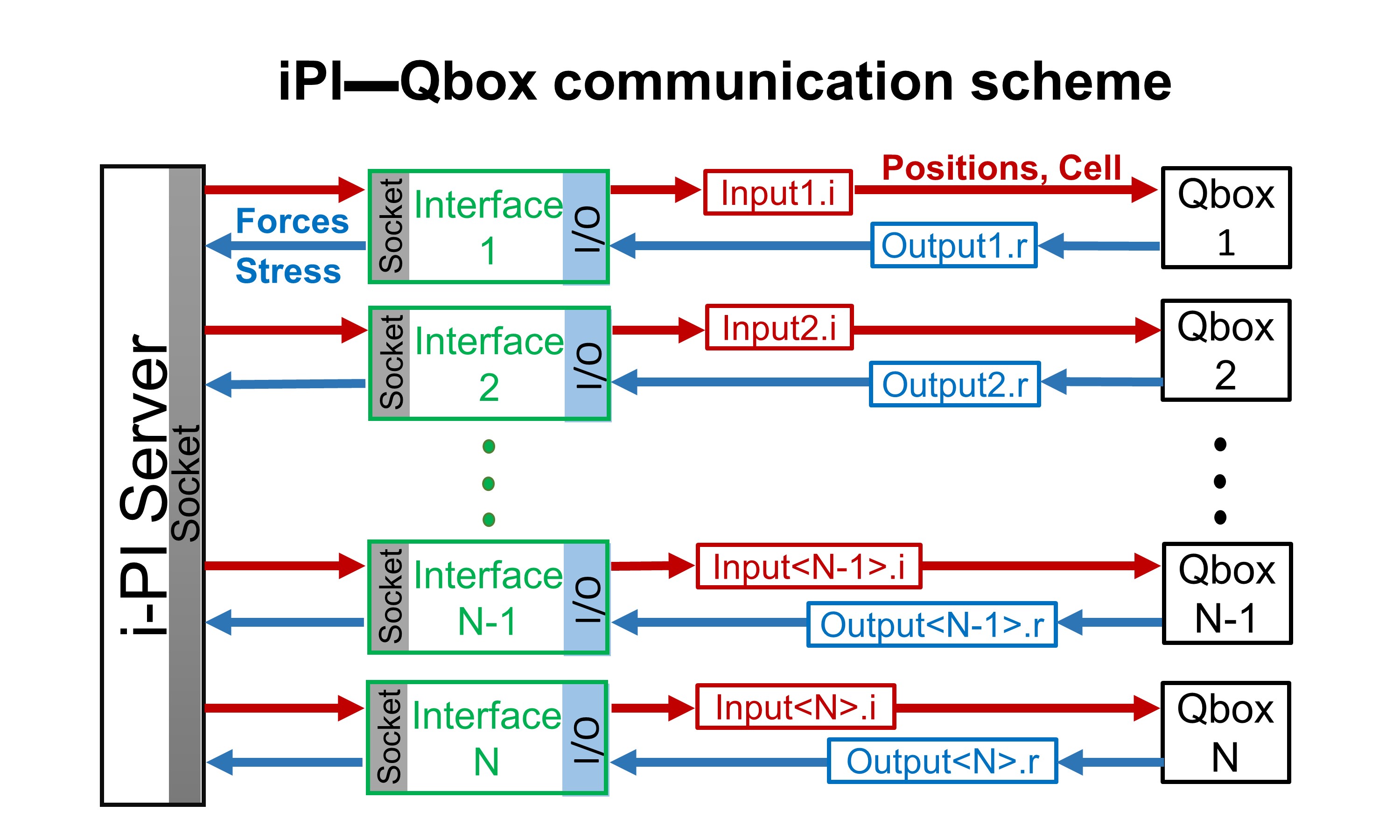}
    \caption{Communication scheme between the i-PI server and Qbox servers via interface clients.}
    \label{fig:interface_comm}
\end{figure}

We coupled the path integral package i-PI \cite{ipi_Kapil_2018} with the first principles molecular dynamics (FPMD) code Qbox \cite{Qbox_Gygi_2008} following a client-server protocol. In an iPI---Qbox coupled simulation, i-PI moves ions while Qbox acts as the DFT engine that supplies forces and stress tensors from first-principles DFT calculation. We implemented the coupling by developing an external python interface client that maintains constant communication between an i-PI server and several Qbox servers (see Fig \ref{fig:interface_comm} for the communication scheme). 

The i-PI code is designed to keep communication with its clients via Unix/internet sockets. Qbox has input/output (I/O) file-based client-server mode where an external client program can write instructions to an input file, which Qbox executes. An output file is written and Qbox awaits for further instruction from the client. By utilizing Qbox's client-server mode, we designed an iPI---Qbox interface that acts as a client for both i-PI and Qbox servers. 

The functioning of the interface between Qbox and iPI can be understood with an example of a path integral molecular dynamics simulation with N beads where for each bead, one instance of Qbox is used (though it is possible to use  a smaller number of Qbox instances than beads). In this example, one launches an interface client for each Qbox server instance.
During a MD move, i-PI updates the coordinates and cell parameters (for NPT simulations) for each bead and sends them through sockets to the respective interface clients. After receiving such information, each interface client prepares an input file with the necessary commands for its Qbox server. After Qbox receives this input, it executes the commands sequentially and upon completion it creates an output file, which contains the resulting forces and stress tensors. Then, the interface client reads the output file and collects the forces and stress tensors, and sends them to the i-PI server through a socket. Using this information, i-PI updates the ion coordinates and the cell-parameters (if required) again and sends the new coordinates and cell parameters to the interface clients.

\section{Methods}

We coupled Path integral (PI) FPMD simulations with a colored noise generalized Langevin equation (GLE) to reduce the number of replicas required in PI simulations, while reproducing quantum fluctuations of the ionic positions \cite{QT_Ceriotti_PRL_2009,QT_Ceriotti_JCTC_2010}. In particular,  we used  a refined version of the colored noise thermostat, named PIGLET \cite{PIGLET_Ceriotti_PRL_2012}, specifically designed to describe systems with strong anharmonic potential energy surfaces.
In addition, we performed FPMD simulations with a single replica and colored noise GLE (a so-called quantum thermostat (QT) \cite{QT_Ceriotti_PRL_2009}).
We note that straightforward path-integral simulations \cite{PI_Berne_Rev_1986,PI_Marx_Rev_1996,PI_hererro_rev_2014} are computationally  demanding, when used in conjunction with forces computed from first principles DFT calculations, especially at low temperatures and for the large supercells required to study disordered or defective systems. 
 In our study, we coupled the FPMD code Qbox \cite{Qbox_Gygi_2008} with the path integral package i-PI \cite{ipi_Kapil_2018} in client server mode, through a python interface (see previous section). In particular, i-PI moves the ions using the forces (and stress-tensors if required) obtained from Qbox. 
 Forces were computed at the PBE level of theory \cite{PBE_Perdew_PRL_1996_1,PBE_Perdew_PRL_1996_2}, using a plane wave basis set with a cutoff of 50 Ry, and optimized norm-conserving Vanderbilt pseudopotentials \cite{ONCV_2015}. Unless otherwise specified, for the PIGLET simulations, we used 16 and 8 beads at 100 K and 250 K, respectively, and 6 beads only for higher temperatures 
 %\footnote{See section S2 in the SI for more information regarding simulations.}.
 We verified that this number of beads provided fully converged results. In our FPMD simulations without NQE (termed as classical) we used a stochastic velocity rescaling thermostat \cite{BDP_JCP_2007}.

In all simulations reported here, we chose 0.5 fs as the time step for integration and we used the preconditioned steepest descent algorithm (PSDA) for the wave function dynamics. Unless otherwise mentioned, we equilibrated the system for at least 2.5 ps, followed by  17.5 ps simulations at 100 K and 12.5 ps simulations at higher temperatures. For the 2-beads NPT PIGLET simulations for the C\textsubscript{216} diamond supercell, we used only 10 ps production run at all temperatures. For C\textsubscript{64} supercell of diamond, we used 22.5 ps production simulations at all temperatures and for all MD (PIGLET, QT, Classical) simulations. In the case of C\textsubscript{512} diamond supercell (at 1000 K), we  performed a 6 ps FPMD simulation, including 2 ps for the equilibration.  

\begin{figure}[H]
\centering
\includegraphics[scale=0.15]{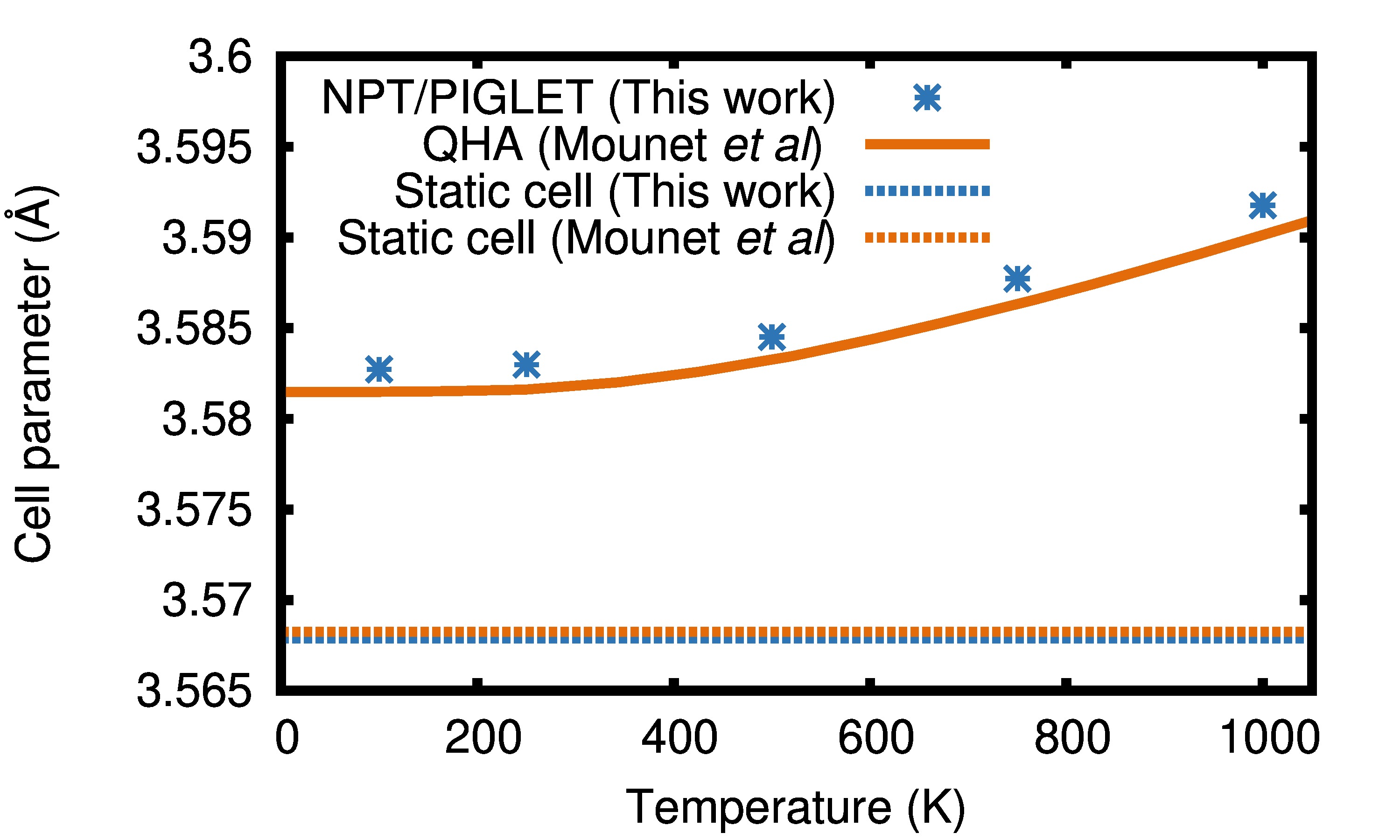}
\caption{Variation of the cell-parameter computed at the PBE level of theory as a function of \textit{T} obtained from 2-beads PIGLET simulations in isothermal-isobaric (NPT) ensemble, compared with previous ``PBE" studies by Mounet \textit{et al.} using the quasi harmonic approximation (QHA) \cite{QHA_Marzari_2005}. Note that our computed static cell parameter is consistent with previous PBE studies of diamond \cite{QHA_Marzari_2005}, and the  thermal expansion of the lattice parameter obtained here including anharmonic effects is in agreement with the quasi-harmonic approximation (QHA) results of Ref. \cite{QHA_Marzari_2005}}
\label{fig:dia_cell_vs_T}
\end{figure}

In our  NVT simulations of diamond we used the static cell parameter obtained by optimizing the stress tensor at the minimum of the PES. In PIGLET simulations, we used an isotropic barostat with a target pressure of 1 atm. Fig. \ref{fig:dia_cell_vs_T}, shows a comparison between the static cell parameter (3.568 \AA) and the cell parameters obtained from the NPT simulations. 

\section{Computational requirements}

Our QT simulations were on average  $\sim$3 times as costly as classical FPMD simulations though, in principle, the two simulations are expected to (and can) have a comparable computational cost. The reason for this overhead, which we think can be greatly improved in future implementations, is that while performing an iPI---Qbox coupled simulation, after an update of the ionic coordinates, a new electronic SCF calculation starts only with a guess of the converged wavefunction of the previous ionic configuration, whereas during a classical FPMD with  Qbox, a wavefunction extrapolation scheme is used by Qbox and consequently, the electronic SCF calculation starts with a better guess. 2-beads PIGLET simulations, as expected, were  $\sim$2 times as costly as QT simulations.

We note that for large cells and/or expensive functionals,   even doubling the required computer time by a factor of 2 may be prohibitive and thus  QT simulations are a recommended, pragmatic choice.

\section{Extrapolation of finite \textit{T} MD bandgaps to 0 K}
%According to electron-phonon renormalization 

In order to extrapolate the finite temperature bands gaps to 0 K, we used the Vi\~{n}a model \cite{Vina_PRB_1984}:
\begin{equation} \label{eq:vina_eq}
 E\textsubscript{gap}(T)=E\textsubscript{gap}(0)-A[1+2/(e^{B/T}-1)], 
\end{equation}
to fit our simulated badgaps which include nuclear quantum effects (NQE).
For results obtained with FPMD simulations, we used a linear extrapolation .

\section{Frozen phonon harmonic (FPH) calculations}
In our  FPH calculations, we used 0.005 a.u. Cartesian ionic displacements to generate the initial dynamical matrix and phonon eigenvectors using the finite difference (FD) module of i-PI. Then we displaced the ionic coordinates along those phonon eigenvectors using the energy scaled normal-mode displacement (ENMFD) module of i-PI with a target energy displacement of 0.001 a.u. During ENMFD calculations, we calculated and saved the Born-Oppenheimer total energy, Kohn-Sham eigenvalues as well the wavefunctions of selected states (e.g., for diamond: three highest occupied and six lowest unoccupied states) for each displaced geometry. We post-processed these saved data to calculate the electron-phonon bandgap renormalizations at various temperatures. We developed a python package named pyEPFD for such post-processing.

Within the frozen phonon method, the temperature dependent, electron-phonon renormalized \(\lambda\)-th Kohn-Sham eigenvalue \((\varepsilon_\lambda\)) is given by \cite{FPH_Bester_2013,Antonius_PRB_2015, FPH_Bester_2016,Monserrat_Rev_2018,FPH_Bester_2019}:
\begin{equation} \label{eq:renorm}
      \varepsilon_\lambda(T) = \varepsilon_\lambda^0 + \sum_{\nu}\frac{1}{2\omega_\nu}\eval{\pdv[2]{\varepsilon_\nu}{u_\nu}}_{u_\nu=0}\qty[n_\nu(T)+\frac{1}{2}],
\end{equation}
where \(\varepsilon_\lambda^0\) is the eigenvalue at the equilibrium geometry and \(n_\nu(T) = (\text{exp}[\omega_\nu/T]-1)^{-1}\), is the Bose-Einstein distribution. The vibrational frequency of the \(\nu\)-th phonon mode \(\omega_\nu\) can be calculated from the second derivative of the Born-Oppenheimer energy \((E_{\text{BO}})\) with respect to the corresponding phonon eigenvector,
\begin{equation} \label{eq:omega}
    \omega_\nu = \sqrt{\eval{\pdv[2]{E_\text{BO}}{u_\nu}}_{u_\nu=0}},
\end{equation}
where \(u_\nu\) is the displacement along the \(\nu\)-th phonon mode.

The pyEPFD package calculates the second derivatives of the Born-Oppenheimer energy \((E_{\text{BO}})\) and the Kohn-Sham eigenvalues \((\varepsilon_\lambda\)) with respect to each phonon eigenvectors by employing the central difference formula,
\begin{equation} \label{eq:central_diff}
    \eval{\pdv[2]{y}{x}}_{x=x_0}=  \frac{y(x_0+ h)+y(x_0-h)-2y(x_0)}{h^2},
\end{equation}
where \textit{h} represents a small change. Using these second derivatives, pyEPFD calculates the renormalized Kohn-Sham eigenvalues utilizing Eqs. \ref{eq:renorm} and \ref{eq:omega}. 

Depending on the displacement chosen for each phonon mode, orbital degeneracies may be lifted in the displaced structure. In such cases, while calculating 
\( \eval{\pdv[2]{\varepsilon_\nu}{u_\nu}}_{u_\nu=0} \)
 applying Eq. \ref{eq:central_diff}, it is necessary to ensure that the electronic eigenvalues used for the positive and negative displacements correspond to the same orbital. Following earlier works \cite{FPH_Bester_2013,FPH_Bester_2019}, we ensured the correct correspondence by projecting wavefunctions with negative displacements on wavefunctions with positive displacements.  

The ionic displacements, Born-Oppenheimer energies, Kohn-Sham eigenvalues, and wavefunctions are obtained from the outputs of ENMFD calculations performed using iPI---Qbox coupling; the pyEPFD package only utilizes them to calculate renormalized Kohn-Sham eigenvalues at a particular temperature. 

\section{Vibrational densities along eigenmodes}

Let \((Q_\nu)\) be the \(\nu\)-th eigenmode expressed in the units of average phonon displacement, \(1/\sqrt{\omega_\nu}\), where \(\omega_\nu\) is the frequency (in atomic units) of the simple harmonic oscillator. The ground state vibrational wavefunction of this harmonic oscillator is,

\begin{equation} \label{eq:vib_wf}
    \phi_\nu(Q_\nu) = \qty(\frac{1}{\pi})^{1/4}e^{-Q_{\nu}^2/2},\,
\end{equation}
and the ground state vibrational density is
\begin{equation} \label{eq:vib_density}
    \abs{{\phi_\nu(Q_\nu)}}^2 =  \qty(\frac{1}{\pi})^{1/2}e^{-Q_{\nu}^2}.\,
\end{equation}
Therefore the ground state vibrational density of a harmonic oscillator is a Gaussian function.  In presence of cubic anharmonicity, i.e., in the presence of a cubic term in the Taylor series expansion of the PES arround \(Q_\nu=0\),  the symmetry of the PES with respect to \(Q_\nu=0\) is broken. Consequently, the ground state wave function  is no longer  a Gaussian due to skewness.  

\begin{figure}[H]
    \centering
    \includegraphics[scale=0.3]{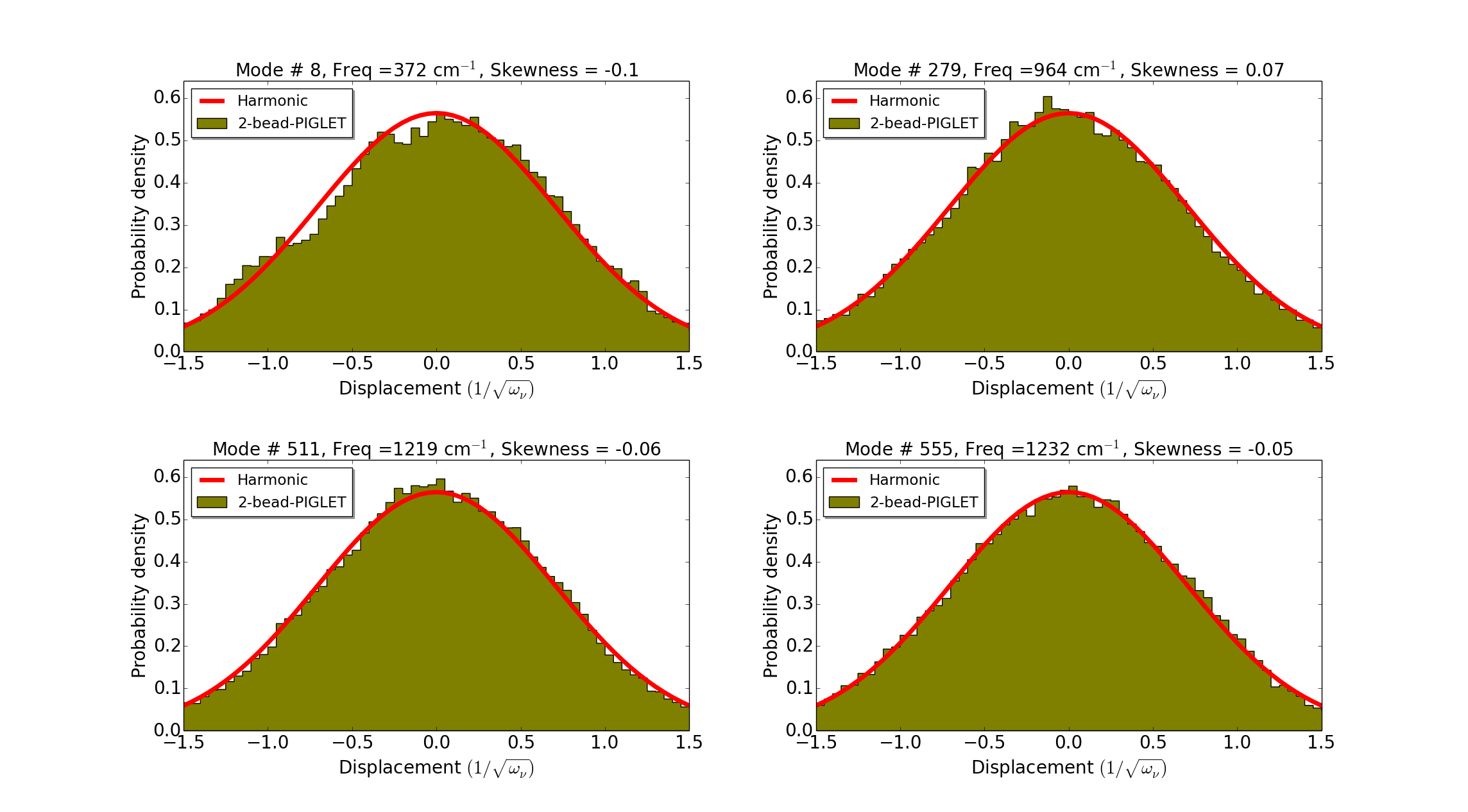}
    \caption{Probability distributions along a few chosen eigenmodes obtained from a 2-bead NVT PIGLET simulation of a diamond C\textsubscript{216} supercell at 100 K are compared with the harmonic density. }
    \label{fig:eigenmode_dist}
\end{figure}

Within the pyEPFD package, we developed a tool named xyz2nmode, that projects xyz coordinates obtained from MD simulations onto eigenmodes \((Q_\nu)\) and returns the projection coefficients.  Using these coefficients, the xyz2nmode program calculates the probability densities along all eigenmodes and furthermore performs statistical test of significance with a null hypothesis  verifying if the obtained probability distribution exhibits zero skewness \cite{skewtest2}. For the 2-bead PIGLET simulation of a C\textsubscript{216} diamond supercell at 100 K, the null hypothesis was rejected within 99\% confidence interval for 234 eigenmodes out of 645 active eigenmodes. In other words, with 99\% confidence we can infer that for 234 eigenmodes, the probability distribution deviates from a Gaussian distribution due to skewness; a few of these modes are shown in Fig. \ref{fig:eigenmode_dist}.  For none of these modes, the Fisher-Pearson coefficient of skewness \cite{CRC_stat_book} is larger than 0.12, indicating that the cubic anharmonicity along those modes are not strong. Nevertheless, the weak anharmonicity found here  is sufficient to break the symmetry of the crystal, thus lifting the degeneracy of the VBM and CBM of crystalline diamond.  

\section{Orientational tetrahedral order parameter for diamond and amorphous Carbon}
\begin{figure}
    \centering
    \includegraphics[scale=0.24]{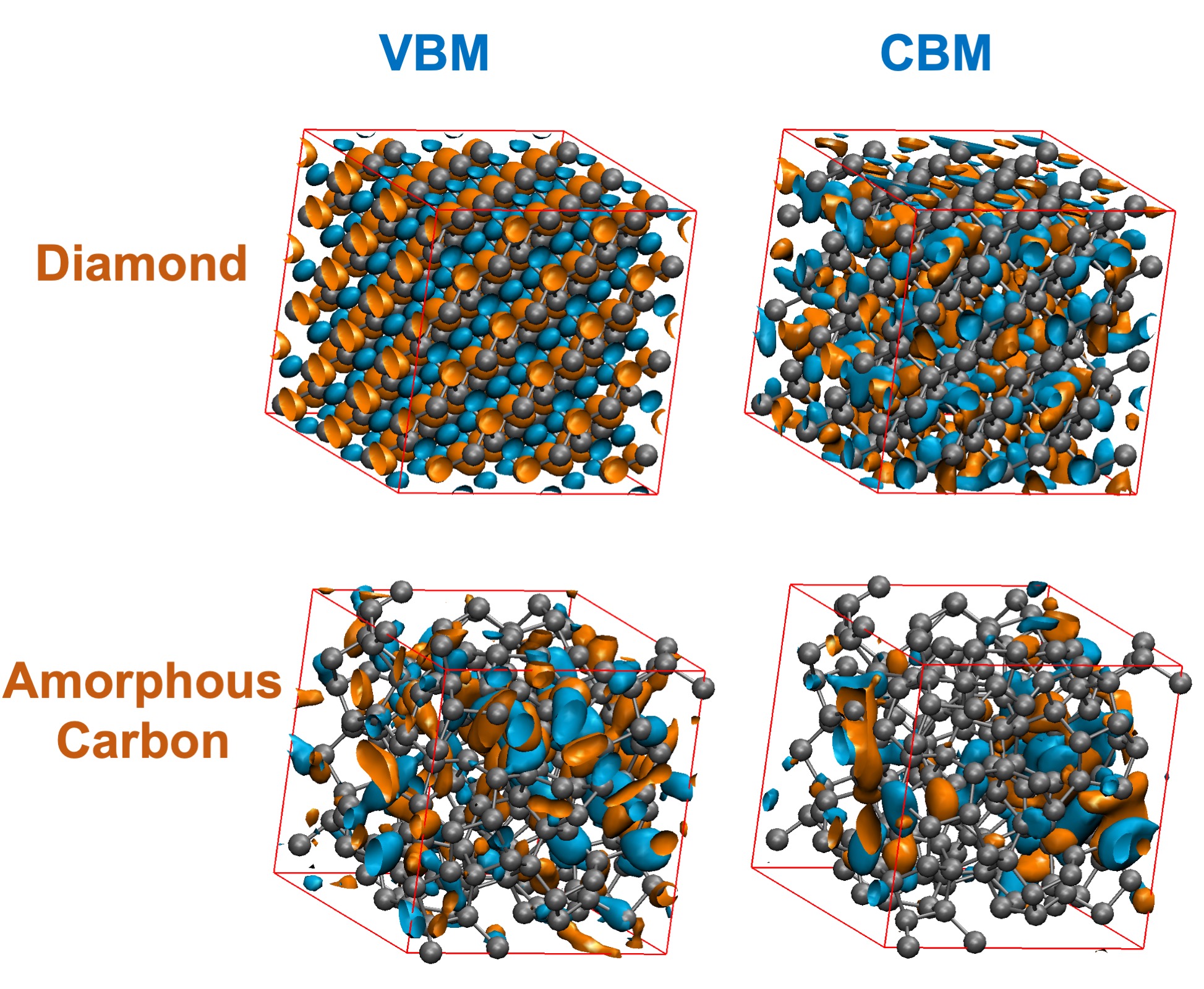}
    \caption{Wave functionas of valence band maximum (VBM, left column) and conduction band minimum (CBM, right column) for 216 C-atom sample of diamond (top row) and amorphous Carbon (bottom row) at equilibrium geometry. For diamond only one of the three degenerate VBM and 6 degenerate CBM states are shown, while for amorphous Carbon those states are not degenerate. }
    \label{fig:vbm-cbm-wf-plot}
\end{figure}

To characterize the local symmetry breaking in crystalline and amorphous carbon, using a python tool named Order \cite{order_code}, we calculated the orientational tetrahedral order parameter \cite{oto_ref_mol_phys_1998, oto_ref_Errington_Nat_2001},
\begin{equation} \label{eq:oto}
    q =  1 - \sum_{j=1}^{3}\sum_{k=j+1}^{4}\qty(\cos{\theta_{jk}} +\frac{1}{3})^2,\,
\end{equation}
where \(\theta_{jk}\) is the angle formed by a given Carbon atom and its {\it j-}th and {\it k-} th neighbors. At the equilibrium geometry of the diamond crystal, six C---C---C angles are equal for all C-centers, and therefore the order parameter for all of these C-centers is 1. 
%Note, when the crystal begins to vibrate, the C---C bond length altering vibrations (such as stretching modes) would still preserve the order parameter since it does not depend on bond lengths and therefore, translationally invarient. 
However, when all six C---C---C angles are not equal, i.e., a local symmetry breaking occurs, then the order parameter for that C-center is less than 1. Such local symmetry breaking  lifts the degeneracies of VBM and CBM of diamond; note that the VBM and CBM those states are  delocalized over the whole crystal (see Fig. \ref{fig:vbm-cbm-wf-plot}) and respond differently to any local breaking of symmetry and deviation from the equilibrium geometry. 

\begin{figure}
    \centering
    \includegraphics[scale=0.2]{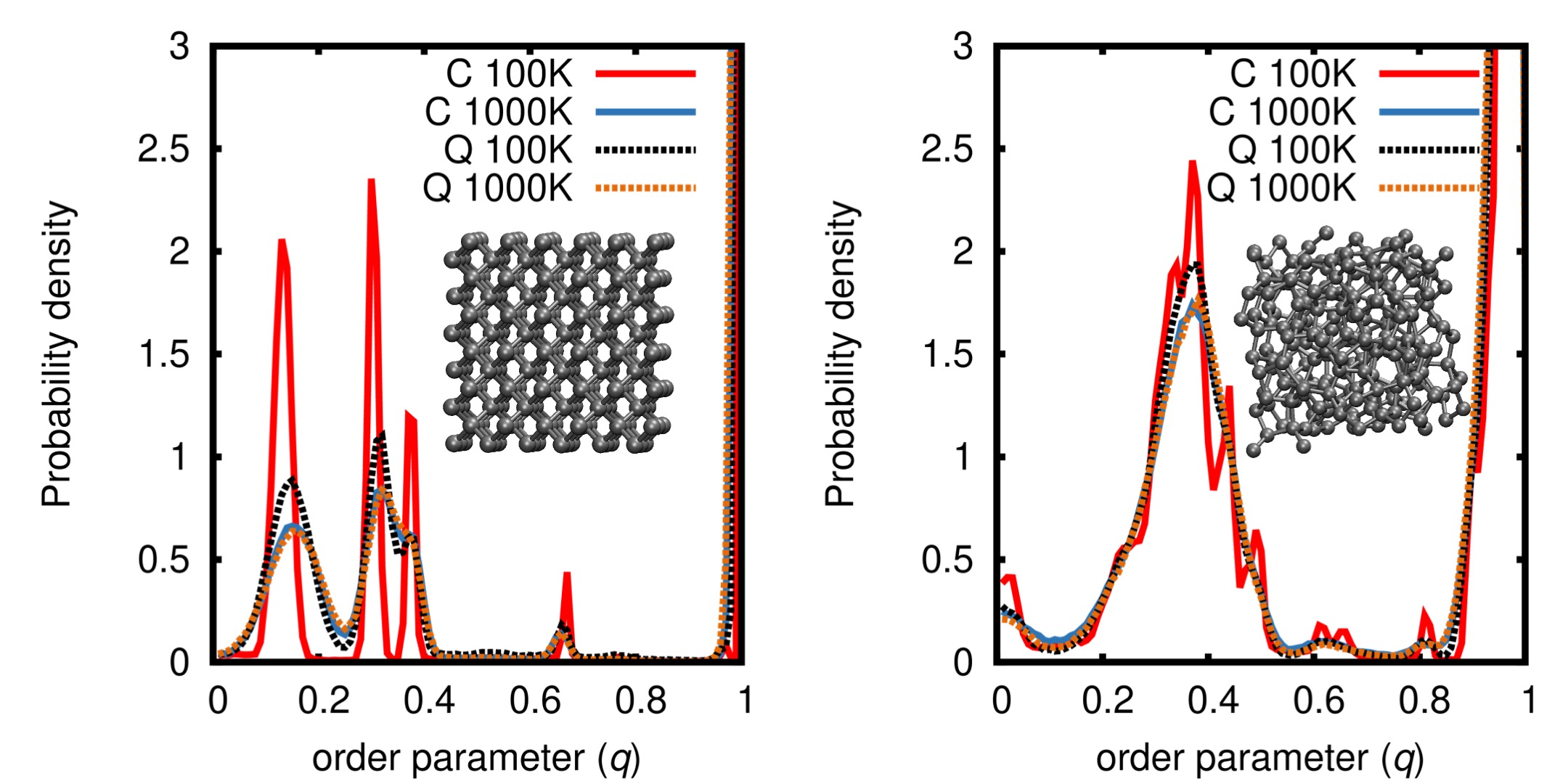}
    \caption{The probability distribution of orientational tetrahedral order parameter computed using Eq. \ref{eq:oto} from the MD trajectories with (quantum nuclei labelled as Q) and without nuclear quantum effect (classical nuclei, labelled as C) at 100 K and 1000 K . Left panel shows the results for C\textsubscript{216} diamond supercell where 2-bead PIGLET approach is used to describe nuclear quantum effect. Right panel shows the results for amorphous-C\textsubscript{216} sample where nuclear quantum effects are included with 16-bead and 6-bead PIGLET approach at 100 K and 1000 K, respectively. }
    \label{fig:oto_dia_aC}
\end{figure}

Figure \ref{fig:oto_dia_aC} shows the order-parameter probability distribution for diamond and amorphous Carbon obtained from MD simulations with (labelled as Quantum, Q) and without (labelled as classical, C) NQE. In both cases, there is a peak around \(q = 1\) indicating that the local tetrahedral orientational order is preserved for the majority of the C-centers. For classical MD trajectories of diamond at 100 K, we observe 4 additional distinct peaks at the interval \(0 < q < 0.9\) due to the local variation of the bending angle. For the disordered amorphous C, however, we only see a very broad peak in that region, which is not significantly changing either with temperature or MD sampling (classical versus quantum) used, i.e., the peak is not sensitive to the amplitude of the oscillations. Therefore. broadening of these multiple sharp peaks can be attributed to local orientational disorder. Note, due to such disorder, the VBM and CBM of the amorphous C sample are not degenerate even at the equilibrium geometry.

For diamond at 100 K, compared to classical MD, we observed significant broadening of the order-parameter probability distribution in the interval \(0 < q < 0.9\) when NQE is included. This shows that local disorder is present in diamond even at very low T due to large amplitude quantum vibrations and consequently, the degeneracies of the VBM and CBM are  lifted.  

\newpage
\section{Additional Figures}
\begin{figure}[H]
\centering
\includegraphics[scale=0.15]{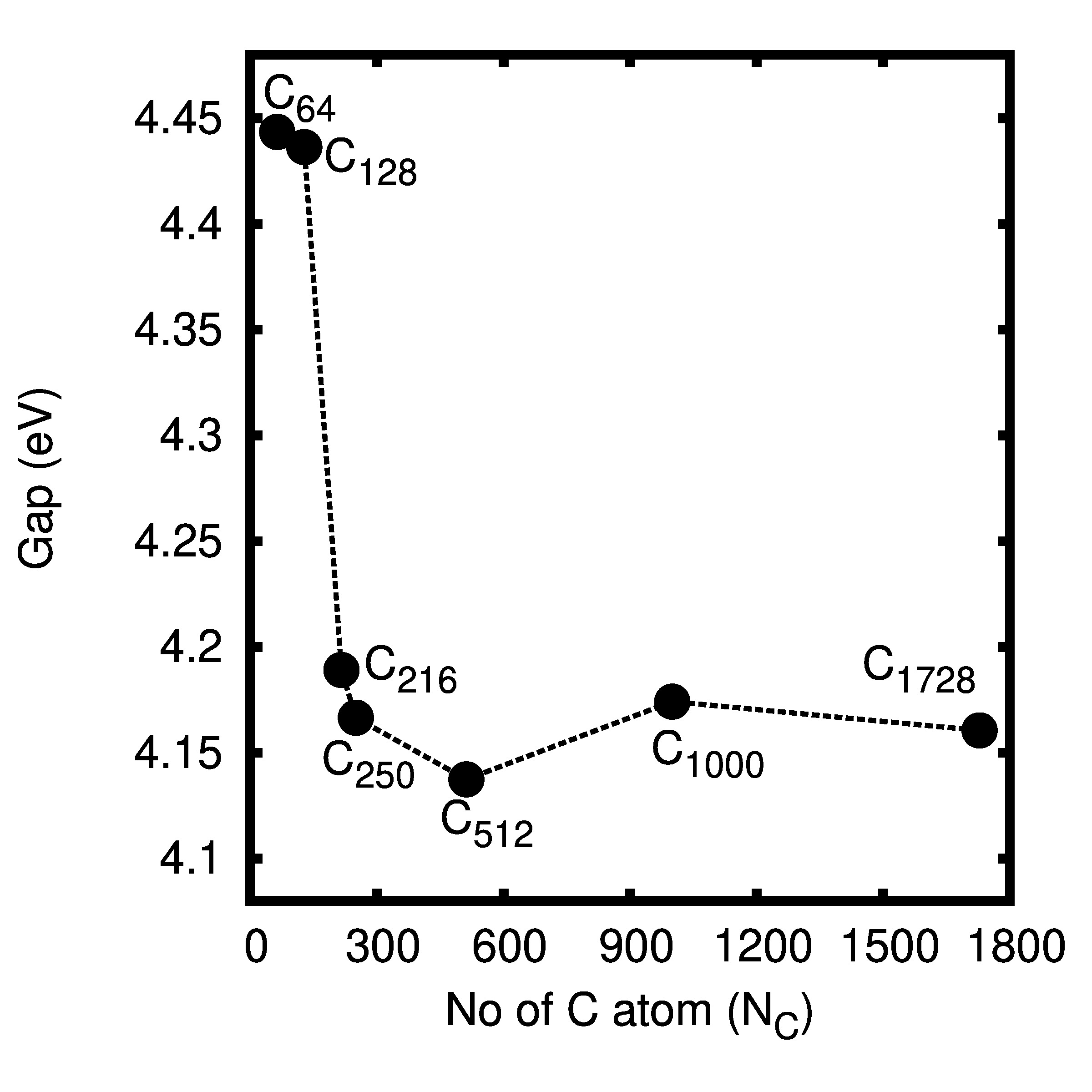}
\caption{Variation of the indirect band gap of diamond as a function of supercell size.}
\label{fig:dia_gap_vs_supercell}
\end{figure}

\newpage

\newpage
\begin{figure}[H]
\centering
\includegraphics[scale=0.2]{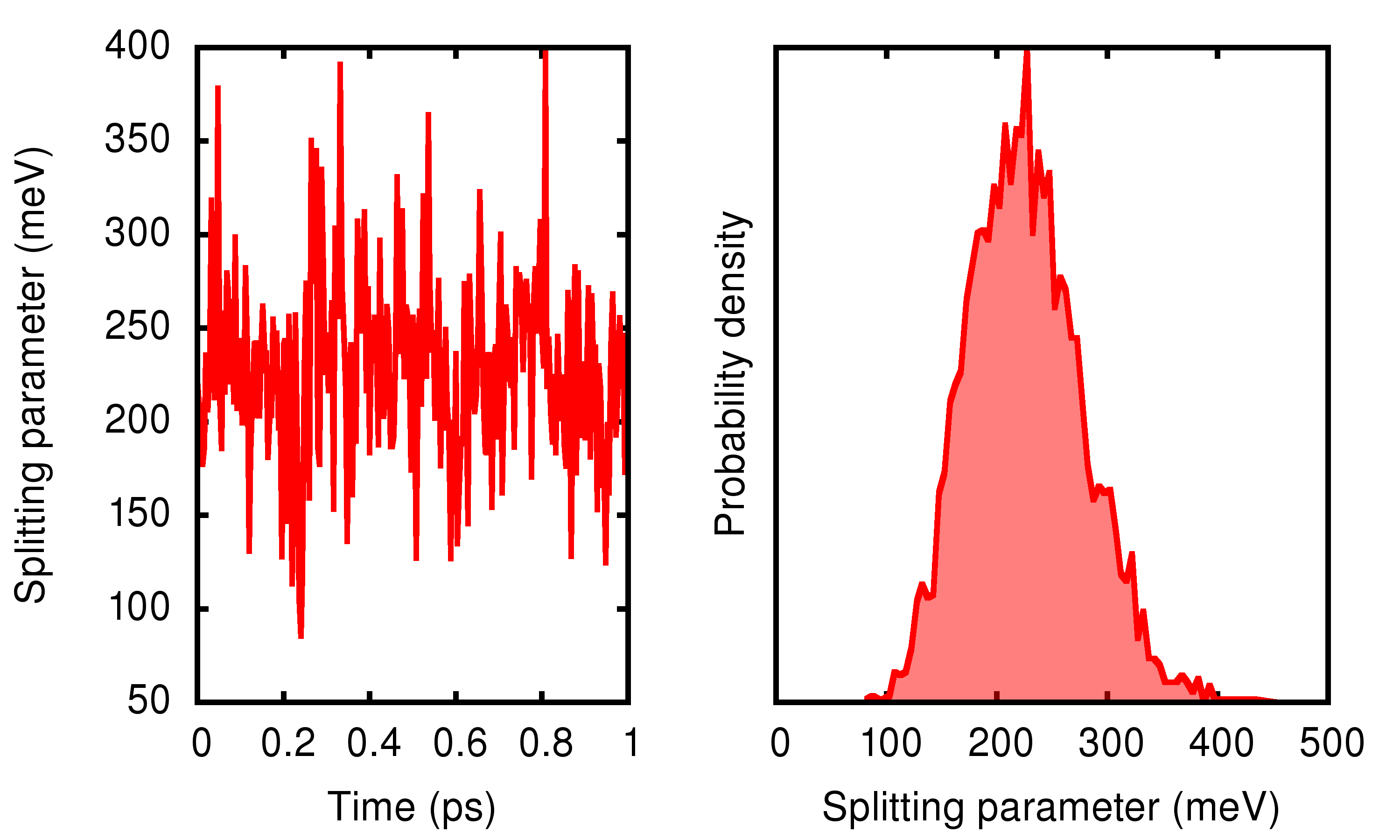}
\caption{Left:  The time evolution of the difference between center and edge gaps (the splitting parameter) as a function of time, computed for a C\textsubscript{250} supercell with 2-beads PIGLET simulation at 100 K. Right: The distribution of the splitting parameter over the full production run (17.5 ps) for the same supercell, obtained from sampling over 4375  configurations.}
\label{fig:penta_gap_compare}
\end{figure}

\newpage
\begin{figure}[H]
\centering
\includegraphics[scale=0.25]{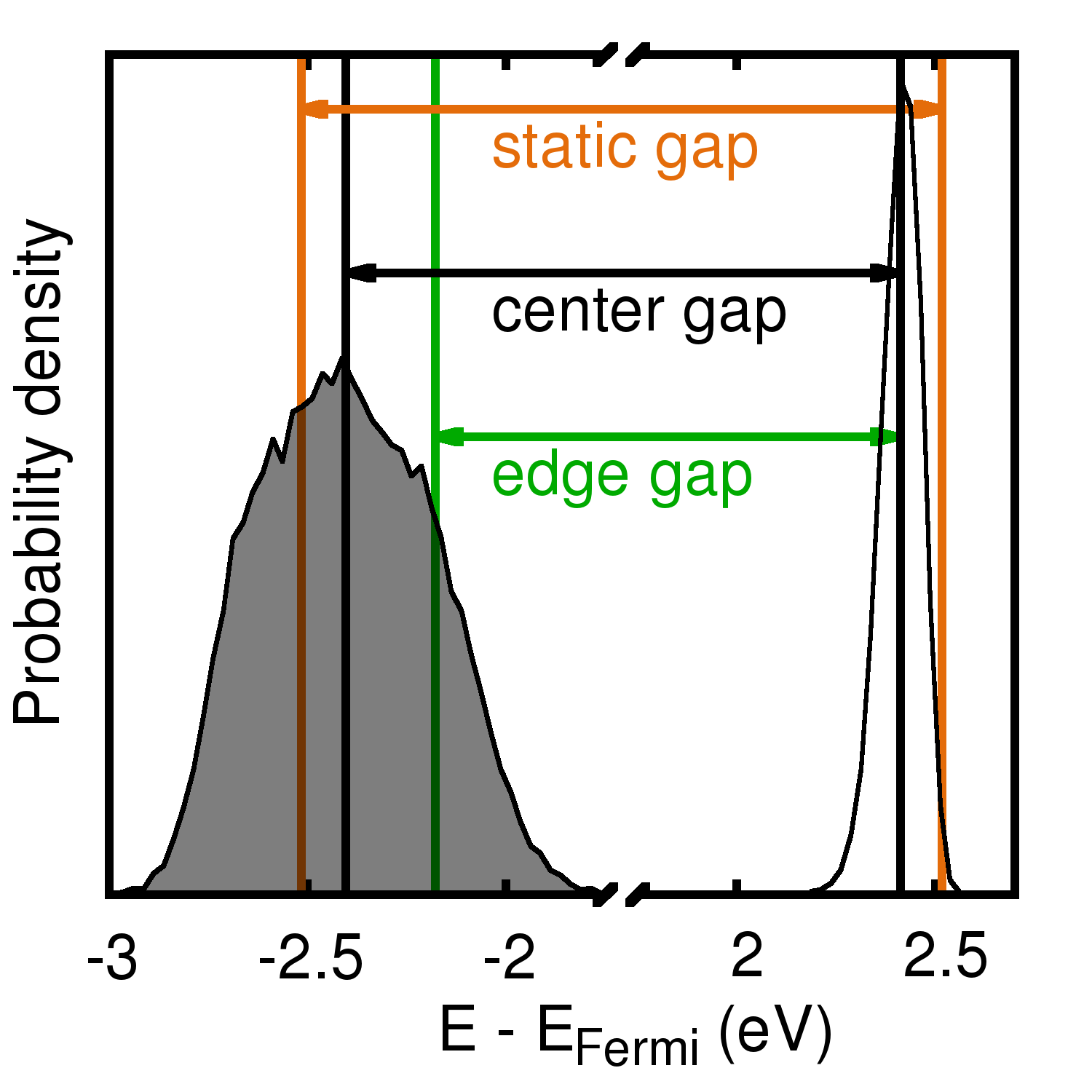}
\caption{Electronic density of states (EDOS) in close proximity of the  HOMO (shaded) and LUMO for an isolated pentamantane molecule, C\textsubscript{26}H\textsubscript{32} (T\textsubscript{d}), obtained from a 16 beads PIGLET simulation performed at 100 K. At the equilibrium geometry, HOMO and LUMO are 3-fold and singly degenerate, respectively. 
The green vertical line represents the thermal average of only the highest eigenvalue among the occupied molecular orbitals. 
The black vertical lines are the thermal averages of the centers of the state distribution. The orange vertical lines represent the HOMO and LUMO energies at the equilibrium geometry.}
\label{fig:penta_homo_lumo_edos}
\end{figure}

\begin{figure}[H]
\centering
\includegraphics[scale=0.15]{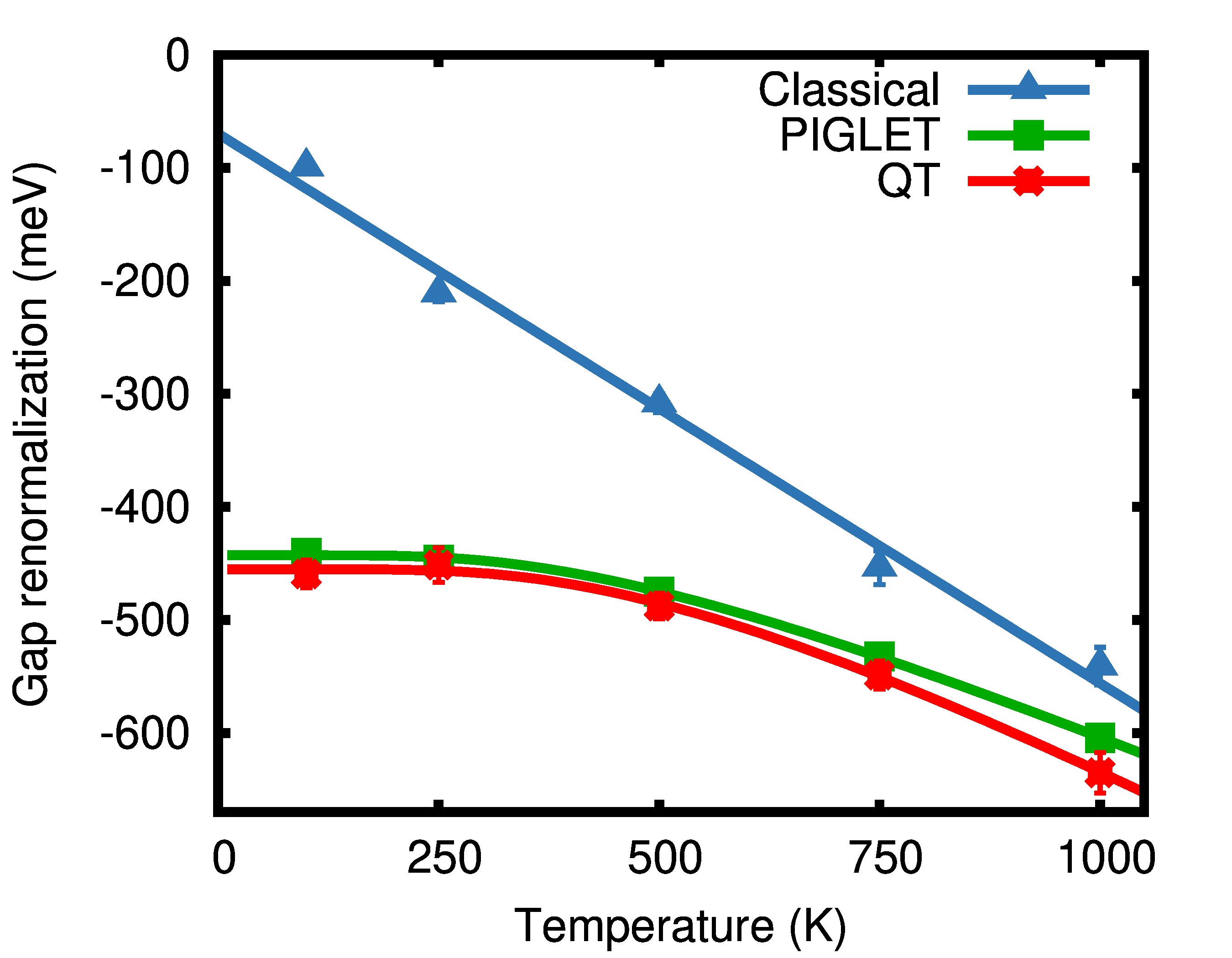}
\caption{Renormalization of the fundamental gap of a pentamantane molecule,  C\textsubscript{26}H\textsubscript{32} (T\textsubscript{d}), obtained from Classical, PIGLET, and QT simulations. The renormalizations are reported relative to the edges of the HOMO and LUMO distributions.}
\label{fig:penta_gap_compare}
\end{figure}

\newpage
\bibliographystyle{apsrev4-2}
\bibliography{manuscript1.bib}% Produces the bibliography via BibTeX.